\documentclass[12pt]{article}
\usepackage{bbm}

\newcommand{\expec}[1]{< #1 >}             % exp. value
\newcommand{\eqref}[1]{(\ref{ #1 })}
\newcommand{\betr}[1]{\left| #1 \right|}     % modulus
\newcommand{\cc}[1]{\overline{#1}}           % complex conjugation
\newcommand{\im}{Im}                % real part
\newcommand{\re}{Re}                % imaginary part
\newcommand{\Z}{{Z}}
\newcommand{\y}{\hat{y}}              % the quantum auxillary field y
\newcommand{\yRI}{\hat{y}^{R,I}}
\newcommand{\py}{{\hat{\pi}^{(y)}}}     % its conjugate momentum
\newcommand{\pyRI}{\hat{\pi}^{(y)R,I}}
\newcommand{\pyI}{\hat{\pi}^{(y)I}}

\newcommand{\pf}{\hat{\pi}}           % its conjugate momentum
\newcommand{\fluc}[1]{(\Delta #1)^2}      % square of fluctuations
 
\newcommand{\cre}{\hat{a}^{\dagger}}    % creation
\newcommand{\creR}{\hat{a}^{R\dagger}}
\newcommand{\creI}{\hat{a}^{I\dagger}}
\newcommand{\creRI}{\hat{a}^{R,I\dagger}}
\newcommand{\ann}{\hat{a}}            % and annihilation operators

\def\be{\nopagebreak[3]\begin{equation}} \def\ee{\end{equation}}
\def\ba{\nopagebreak[3]\begin{eqnarray}} \def\ea{\end{eqnarray}}
  
 \def\a{\alpha}

\newcommand{\teta}{\rlap{\lower2ex\hbox{$\,\tilde{}$}}\eta{}}
 
\newcounter{mnotecount}[section]

\begin{document}

\title{On the quantum origin of the seeds of cosmic structure}

\author{Alejandro Perez$^{1,2}$\thanks{perez@gravity.psu.edu}, Hanno
  Sahlmann$^{1,4}$\thanks{h.sahlmann@phys.uu.nl},
  and Daniel Sudarsky$^{1,3}$\thanks{sudarsky@nuclecu.unam.mx}\\[.5cm]
  1. Institute for Gravitational Physics and Geometry,\\
  Penn State University\\
  2. Centre de Physique Theorique, Universit{\'e} de Marseille\\
  3. Instituto de Ciencias Nucleares\\
  Universidad Nacional Aut\'onoma de M\'exico\\
  4. Spinoza Institute, Universiteit Utrecht\\
}

\maketitle

\begin{abstract}
  The current understanding of the quantum origin of cosmic structure
  is discussed critically.  We point out that in the existing
  treatments a transition from a symmetric quantum state to an
  (essentially classical) non-symmetric state is implicitly assumed, but
  not specified or analyzed in any detail. In facing the issue we
  are led to conclude that new physics is required to explain the
  apparent predictive power of the usual schemes. Furthermore we show
  that the novel way of looking at the relevant issues opens new windows
  from where relevant information might be extracted regarding
  cosmological issues and perhaps even clues about aspects of quantum
  gravity.
\end{abstract}

%-------------------------------------------------------------------

\section{Introduction}

%-------------------------------------------------------------------

The origin and evolution of the large scale structure of the
universe constitutes a central aspect of cosmology.  The
widespread view is that our understanding of this aspect has
developed dramatically in recent times, both through the
collection of high precision data in observational cosmology
\cite{Boomerang-Maxima-WMAP}, and through better theoretical
understanding \cite{Guth, Hawking}: The observations are in very good
agreement with the theoretical predictions based on the scale
invariant (Harrison-Zeldovich) spectrum of seed fluctuations and
their subsequent evolution. Such a scale free spectrum would be
hard to explain in standard cosmology because physical processes
in the very early universe that might produce it, would have to act
on scales larger that the Hubble radius.This problem is absent
if one considers inflationary cosmology, which seems needed as
well in order to deal with well known puzzles arising in standard
cosmology \cite{MotivInflation}.

What is more, inflation produces a scale free spectrum of quantum
mechanical fluctuations for the quantized inhomogeneous component of
the inflaton field, and thus seems to give birth to these seeds of
structure through a quantum process.

It seems that we have in fact here one of the big successes of
theoretical physics in recent times (besides the remarkable, and
undisputed achievements that it signifies for observational
cosmology), and these facts are widely viewed as confirmation of an
inflationary stage in the very early universe.\footnote{Note however
that recent observations indicate an anomalous lack of power at large
 angular scales, a fact that has lead to some workers in the field to
start questioning the viability of the inflationary model
\cite{Low-l-anomaly}. See also the alternative explanation in
\cite{Inflation-Wald}.}

In the present paper, we will be concerned with one specific part of
the picture summarized above, namely the suggestion that inflation
not only provides a satisfactory understanding of the evolution of
the inhomogeneities, but even an explanation of their origin: The
observed similarity between the spectral properties of the
(classical) density fluctuations in the early universe and the
quantum mechanical fluctuations of the inflaton field after inflation
indeed suggest much more than a coincidence. However, we want to draw
attention to the fact that a detailed understanding of the process
that leads from the quantum mechanical to the classical fluctuations
is lacking. What is needed to justify the connection between the two
spectra is a mechanism that transforms quantum mechanical
uncertainties into classical density fluctuations.

This issue has in fact been treated by many  authors
%  such as J. Hartle,  A. Staribinski, J. Halliwell, B. Hu,
%R. Parentani,  W. Zurek,
(see  \cite{Starobinski1,Kiefer1, Parentani, Zurek, Decoherence,
Mukhanov,
Castagnino:2002ic,
Halliwell-85,
Correlations,
Hartle-1997}
%,Starobinski2
 for example),
  most of whom seem to suggest that the subject has been clarified and
  thus, at least on a fundamental level, possesses no further mystery.
  Indeed most of the proposals seem to invoke nothing beyond standard
  physics to explain the coincidence of the spectra. On the other
hand this same body of literature can be considered as an
indication that underneath all of the reassurances to the
contrary, there is a remaining degree of discomfort, and certainly
the fact that the different authors point to slightly different
schemes for this quantum to classical transmutation, indicates
that each author does not find the schemes espoused by their
colleagues to be fully satisfactory. Indeed we feel that, at
least on a fundamental or conceptual
  level, the treatments proposed are missing something, and we will
  try to pinpoint the place where there is a missing step in the most
  well known ones. This will lead us to argue that in fact something
beyond standard physics seems to be required if the essential
picture and its successful degree of predictive power is to be
preserved. At the same time we underscore the richness of the
physical information that is being overlooked when the problem is
not addressed head on.

% For a discussion of some of the
%possibilities we refer the reader to section \ref{sec_other}.

To illustrate the physics behind that missing aspect, we introduce
a simple phenomenological model involving a dynamical
collapse of the wave function as the mechanism that leads to the
transition to inhomogeneity. It is phenomenological in the sense
that it does not explain the transition to inhomogeneity by some
particular new physical mechanism, but merely gives a rather
general parametrization of such a transition.  We will briefly
discuss some of the characteristics that the new physics would
have to include if we wanted it to justify our analysis. We also
show how the value of some parameters can be studied by
comparisons with observational data.

%We also investigate a closely related alternative similar in
%spirit to the many world interpretation of quantum mechanics,
%which including novel elements, would also have to be regarded as
%new physics.

Since one is ultimately dealing with connecting quantum mechanical
  quantities to measurement results, one might be tempted to consider
the issue described above as related to old and well known
interpretational problems of Quantum Mechanics, and thus to dismiss
it as inconsequential as far as the physical predictions for
observations are concerned. After all, these interpretational
problems of Quantum Mechanics can be mentioned in every instance
where use of the theory is made, and we have all grown accustomed to
the fact that, in practice, those issues have no incidence whatsoever
on our ability to make correct predictions.
%\footnote{We could think for instance in the
%  situation that arises when we describe the standard EPR setup from
%  the point of view of two different inertial observers. In that case
%  we could also inquire about the time at which the state of the
%  second particle's spin collapses due to our measurement of the first
% particle's spin. We know that Quantum Mechanics provides no answer
% that can be considered globally satisfactory for all points of view.
%  However one can take either answer and make detailed predictions. In
%  the analysis existing in the literature about the birth of the
%  cosmological inhomogeneities we do not seem to have any such
% description available. }
We will argue that quite to the contrary, the situation at hand,
namely the cosmological setting, is such that the way one deals with
these issues could have an impact on the predictions for
observations. Furthermore, the standard rules that one can rely on,
to settle all practical ambiguities in ordinary quantum mechanical
situations, are not available in the present case.  The reasons for
these differences are threefold: First the object that is being
treated quantum mechanically is the entire universe, and thus the
standard separation of the system into subsystem of interest,
observer, and environment, becomes unjustifiable and would be in fact
subject to completely capricious choices. Second, that we have at our
disposal just a single system -- our universe --, so the recourse to
the statistical ensemble interpretation of the result of a
measurement in quantum mechanics is not directly available. And third
the fact that not only do the observations pertain to the very late
causal future of the assumed quantum to classical transition, but in
fact the existence of the observers themselves depends upon the
outcome of the measurements. In our view these facts indicate, not
only, that this aspect of cosmology offers a rather unique
opportunity to focus on these so called ``interpretational aspects of
quantum theory" -- and to look for clues that might lead to a better
understanding of them -- but point in fact to the conclusion that any
satisfactory treatment of it requires a step beyond what we currently
have as established interpretational models of quantum mechanics.  We
will refer to this unknown aspect as new physics.

The article is organized as follows: In Section \ref{sec_standard} we
review the standard description of the inflationary scenario for the
origin of the seeds of cosmic structure. In Section
\ref{sec_critique} we make a general critique of the standard
description. In Section \ref{sec_other} we critically discuss the
main ideas that have been put forward regarding the transition to
classicality. In Section \ref{sec_element} we point out the nature of
the missing element and propose some ideas. In Section
\ref{sec_linear} we review the description of linearized fluctuations
maintaining the framework that seems necessary to make our proposals
in a clear and transparent manner. In Section \ref{sec_quantum} we
proceed to make the quantum mechanical treatment of the field
fluctuations within the setting necessary for implementing our ideas.
In Section \ref{sec_collapse} we show how to implement our proposal
within the standard discussion of the origin of the seeds of cosmic
structure. In Section \ref{sec_observation} we recover the
predictions that must be compared with observational results. Section
\ref{sec_further} contains comments and calculations into some
further directions, and we end in Section \ref{sec_disc} with a
discussion of our conclusions.

%The Appendices are included for completeness: Appendix
%\ref{app_spherical} deals with an alternative derivation of our
%result using from the onset a spherical expansion. Appendix
%\ref{app_newton} gives a simple account of the nature of the
%physical observable in our problem and thus the ``gauge
%invariance" of the quantity we study. Appendix
%\ref{app_alternative} gives a brief description of an alternative
%scheme of collapse which gives results slightly different than the
%one followed in the main text. Appendix \ref{seca5} serves to give
%some details of a relative cumbersome calculation pertaining to
%section \ref{sec_observation}.

Regarding notation we will use signature (-+++) for the metric and
Wald's convention for the Riemann tensor. We will use units where
$c=1$ but will keep the gravitational constant and $\hbar$ explicit
throughout the paper.

%-------------------------------------------------------------------
\section{The standard picture}
\label{sec_standard}
%-------------------------------------------------------------------

As mentioned in the introduction there are many treatments of the
subject differing in technical as well as conceptual aspects.
However, they do share a  ``standard core" which we resume here
for the benefit of the reader, with no ambition toward
completeness or rigor\footnote{The  reader can use  for instance
\cite{branderberger} as an up to date review of the subject}.

1) Start with the action of a scalar field coupled to
gravity.\footnote{That we can assume the universe to be devoid of
matter except for the scalar field is in fact the result of the
early inflationary period that is thought to precede the regime we
are describing.} \be \label{eq_action} S=\int d^4x \sqrt{-g} [
{1\over {16\pi G}} R[g] - 1/2\nabla_a\phi \nabla_b\phi g^{ab} -
V(\phi)], \ee where $g$ is the space-time metric and $\phi$ the
inflaton scalar field.

2) One splits the fields $(g,\phi)$ into a homogeneous-isotropic {\em
background} plus an arbitrary {\em fluctuation} assumed to be small,
i.e. $g=g_0+\delta g$, $\phi=\phi_0+\delta\phi$. The background
geometry $g_0$ is assumed to be the Robertson Walker space-time \be
ds^2= -dt^2 +
  a(t)^2(dx_1^2+ dx_2^2 + dx_3^2),
\label{RW} \ee where $a(t)$ is the scale factor and $t$ is the
cosmological time. Note that for simplicity we have restricted
attention to the case of flat spatial slices ($k=0$ in the usual
notation). The homogeneous and isotropic background field $\phi_0$
satisfies the equation of motion $g_0^{ab}\nabla_a \nabla_b
\phi_0+\partial_{\phi}V[\phi_0]=0$ which due to the symmetry
assumptions becomes \be \label{back} \phi_0^{\prime\prime}+3\
\frac{ a^{\prime}}{a}\ \phi_0^{\prime}+
\partial_{\phi}V[\phi_0]=0,
\ee where primes indicate derivatives with respect to ordinary cosmic
time $t$. The second term acts as a friction term. One assumes the
so-called slow rolling condition $\phi_0^{\prime\prime} =0$ (for an
analogy think of the motion an object falling with terminal velocity
under the influence of gravity in a viscous medium). The
energy-momentum tensor of the background field is thought to be
dominated by the potential term, so one takes it to be approximated
by
\[T^{(0)}_{ab}[\phi_0]=V[\phi_0]g_{ab}.\]
To achieve slow rolling, the potential has to be sufficiently
flat in the region where the initial conditions for $\phi_0$ are
set. Consequently Einstein's equations imply that during the
inflationary regime we have:
\begin{equation}
\label{eq_phi0}
a(t) = A e ^{Ht},
\phi(t)=\phi_0(t)=\phi_{initial}+\phi_0^{\prime} t
\end{equation}
where $H=\sqrt{ (4/3) \pi G V_0} $ and $V_0=V[\phi_{initial}]$,
using the fact that during the regime of interest the potential is
essentially constant.

3) Now, turn to the fluctuations, and concentrate on the matter
sector:\footnote{A more detailed treatment would also involve
discussion of the metric perturbations, and issues of gauge. Note
also that in some more recent treatments \cite{Mukhanov},
quantization is applied directly to a gauge invariant combination
of scalar field and metric perturbations.} The scalar field
$\delta \phi(x)$ can be treated as a standard, scalar field on the
background (\ref{RW}). It is useful to decompose it into Fourier
modes, since  they are uncoupled due to the symmetry of the
background. For simplicity we introduce an infra-red cutoff by
assuming the spatial slices to be $\Sigma:=[0,L]^3$, i.e. a box of
length $L$. Then the total Lagrangian for $\delta \phi$ (derived
from (\ref{eq_action})) can be written as a sum of
  mode Lagrangians
\be
\label{lagra}
{\cal L}_k=\frac{a^3}{2}\left[|{ \delta
    \phi^{\prime}}_k|^2-\frac{k^2}{a^2}|\delta \phi_k|^2\right]
\ee
for the modes
$\delta\phi_k=\int_{L^3}\exp(-ikx)\delta\phi(x)$
(with $k_iL/(2\pi)\in \Z$ ($i=1,2,3$)). During inflation the
equations of motion from (\ref{lagra}) become \be \label{eom}
\delta \phi^{\prime\prime}_{k} +3\ H \ \delta\phi^{\prime}_{k}
-\left(\frac{k}{a}\right)^2 \delta \phi_{k} = 0, \ee where one is
using that Einstein's equations imply $a^{\prime} /a=H$ in that
regime.
% (see eqs. (\ref{eeee}) and (\ref{tttt})).
{  We remind the reader that at this point one is not indicating that
there are inhomogeneities of any definite size in the inflationary
universe, but merely one is considering what
 would be the dynamics of any such small inhomogeneity if it existed.
The issue of their presence and magnitude is dealt with at the
quantum level, where the  fluctuation field is just an operator, and
the above mentioned matter is associated with the state of such
quantum field. In fact, the next step is the quantization} of the
field. Note that in terms of the Fourier modes, the Hilbert space for
the quantum field will be a direct product of Hilbert spaces -- one
for each mode.

4) For quantization, one needs to chose a vacuum state for the field
and assume it represents the initial state at the onset of inflation.
{ This state is required to be homogeneous (i.e. all n-point
functions are invariant under translations and rotations in the
spatial slice). After all, the point is to explain the emergence of
the inhomogeneities, rather than merely to assume their presence at
the onset of inflation and to study their evolution.} There is an
obvious problem in this step, because the choice of a vacuum state is
not unique in space-times that do not possess a time-like Killing
field. Though a difficult issue in principle, in practice the results
do not depend much on the choice of the initial state as long as it
approaches a canonical form for modes $ \delta \phi_k$ with large
$k$. In the literature a sort of ``instantaneous vacuum state" is
most often chosen as the initial state. In the way we presented
things here, this corresponds to the assumption that at the onset of
inflation the wave functions for the mode $ \delta \phi_k$ and its
canonical momentum $ \pi_k$ are Gaussian with spreads \be \label{vac}
(\Delta \delta \phi_k)^2= \frac{1}{2a^2 k},\quad\quad (\Delta
\pi_k)^2=\frac{\hbar a^2 k}{2}. \ee For $k >> a H$ the friction term
in (\ref{eom}) can be neglected and thus the time evolution of the
mode $ \delta \phi_k$ is just that of a harmonic oscillator (with
time dependent mass). Then the above choice just amounts to the modes
starting out in the ground state of a standard harmonic oscillator.

5) Time evolution of the quantum state: As described above, a
mode initially corresponding to $k >> a H$ is assumed to start
out in the ground state. As $a$ grows larger the proper wave
length of the mode will reach the Hubble radius ($k=a H$) and the
friction term in (\ref{eom}) will start dominating the dynamics.
As the wave length grows larger one can approximate the solution
to (\ref{eom}) by an over-damped oscillator and the value of the
field can be assumed to be frozen at its value at Hubble radius
crossing. At that moment $a=k/H$ so that the fluctuations of
$\delta \phi$ and $\pi_k$ can be approximated by
\begin{equation}
\label{eq_fluc} (\Delta \delta \phi_k)^2= \frac{H^2}{2
k^3},\quad\quad (\Delta \pi_k)^2=\frac{\hbar k^3}{2H^2}.
\end{equation}

6) Now the fluctuations (\ref{eq_fluc}) are identified with (or, in a
more careful formulation such as in \cite{Padmanaban}, ``taken as
indicative of") the (classical) spectrum of inhomogeneities. For
example, often the fluctuations of the energy density in co-moving
coordinates, $\delta \rho$, is used as a measure for the classical
inhomogeneities. It is easily verified that $\delta\rho\sim \phi_0'
a^{-3}\pi_k$. Thus one sets
\begin{equation}
 |{\delta\rho_k}|^2\sim a^{-6} (\Delta \pi_k)^2\sim k^{-3}
\end{equation}
at the time the mode $k$ leaves the Hubble horizon. $\delta\rho$ is
understood as a classical quantity.

7) The classical density fluctuations are now evolved through the
end of inflation and into the regime of standard cosmology. It can
be shown that the spectrum of the fluctuations at the time they
exit the Hubble radius is proportional to the spectrum at the time
they re-enter. Thus one has $(\delta\rho_k)_{{\rm enter}}\sim
k^{-3/2}$ which is the scale free Harrison-Zeldovich spectrum. The
classical fluctuations are then evolved further to take into
account the matter dynamics, leading to the famous acoustic peaks.

8) The resulting evolved spectrum is compared with the observations
of the CMB. Furthermore, these departures from homogeneity and
isotropy are thought to be the seeds for the evolution of the
structure that we see in the our universe, and that are in fact
necessary for our own existence.\\[.5cm]
This is in essence the standard account of the process for obtaining
the seeds of cosmic structure formation that forms the basis of the
analysis of cosmological data, and in particular of the CMB
anisotropies.

%-----------------------------------------------------------------------
\section{A critique of the story}
\label{sec_critique}
%----------------------------------------------------------------------

The previous analysis is remarkable: The universe starts in a
homogeneous state and ends up with inhomogeneities that fit the
experimental data. However, as we will argue in the following
sections, this is not fully justifiable by ``standard physics''. We
should point out from the start that many authors do acknowledge the
existence of a gap in our understanding, for instance
\cite{Padmanaban,Liddle}, and some do propose ways to deal with the
issue in more detail (see ex. { \cite{Starobinski1, Kiefer1,
Decoherence, More-Decoherence, Lombardo:2005iz}). There is also a
large literature on issues of quantum mechanics in the context of
cosmology in general (ex. \cite{Zurek, Castagnino:2002ic,
Hartle-2005, Halliwel-89}).} The discussion that follows is addressed
to those colleagues that are not fully aware of the problem, and to
those that believe that there is no problem at all.

The first point that we want to make is that the above analysis can
not be justified through standard quantum mechanical time evolution.
If this is not already obvious from the particulars of step 6, just
note that as already stated above, the initial quantum state is
symmetric\footnote{The vacuum state defined in step 4 is invariant
under the action of the symmetry group of the background. Even though
it can be written as a superposition of states which are not
symmetric by themselves the superposition does not select any
direction and looks the same at every point on the homogeneous
$t=$constant slices. This is completely analogous to the Poincar\'e
invariance of the  standard vacuum on Minkowski space-time.} (i.e.,
homogeneous and isotropic) and the standard time evolution does not
break this symmetry, thus it can not explain the observed
inhomogeneities. { The states obtained  in other natural schemes,
such as the one arising from the ``No Boundary" boundary condition in
quantum cosmology are equally homogeneous and isotropic, as can be
seen explicitly in eq 8.6 and 8.9 in \cite{Halliwell-85}}.

Let us now look at step 6 in more detail: The essence seems to be
that a classical quantity is identified with (the square root of)
a quantum mechanical expectation value. This suggests that one
might be able to view the procedure of step 6 in the framework of
a semiclassical theory, i.e. one in which a part of the system is
described classically, while another part is described quantum
mechanically. A hallmark feature of such type of descriptions is
that classical quantities are coupled to the quantum sector via
expectation values. While not satisfactory as fundamental
theories, they nevertheless often have some validity as effective
theories. However, we note that two things are remarkable about
the prescription of step 6 and set it apart from semiclassical
theories such as semiclassical gravity: a) In step 6 a classical
quantity couples to the \textit{square-root} of the expectation
value of the \textit{square} of its quantum counterpart. b) The
use of Fourier transforms in step 6. The choice involved in a) is
certainly not the simplest possibility, and not the one chosen in
standard semiclassical treatments. Point b) would not be an issue
if the relation were linear, but raises concern due to the choice
made in a). For instance, what if instead of plane waves one uses
spherical harmonics in the mode expansion? The non linear nature
of a) implies that this will affect the result. Note also that
both a) and b) must be required to achieve the result of the
standard view described in the previous section. If we drop any of
the two we do not get the sought-for inhomogeneities. For
instance, identifying the square root of the expectation value of
$\hat\delta\phi^2(x)$ with the classical counterpart (i.e.
dispensing with b)) leads to a result that is independent of $x$,
and thus it is homogeneous. Identifying  the expectation value of
$\phi_k$ with its classical counterpart (i.e. dispensing with a)),
leads to a homogeneous result (zero, in fact) as well.

One quite often reads the following argument to support a) and b):
A simple calculation shows that $\expec{\hat\phi_k^2}$ is equal to
the Fourier transform of the auto-correlation of the two-point
function of $\hat\phi(x)$, and thus quantifies correlations of the
field at different points. The observations are then argued to be
a measurement of these correlations, because one is measuring the
differences in the temperature of the CMB between different
directions in the sky. We do not find this convincing however,
because although WMAP in fact works as a differential thermometer,
it does so in a way that allows one to obtain a map of the
temperatures in the sky, and it is this map which is subjected to
a Fourier analysis from which the observational power spectrum is
read.\footnote{After all, if it were technically possible to
replace the comparison of temperatures at two different points in
the sky with, say, the comparison of the temperature of a single
point in the sky with the temperature of a reference body within
the satellite, one would not expect the resulting temperature map
to change (unless one is prepared to argue the CMB really is a
quantum system in a higly non-classical state).
%
%To make the situation more transparent let us consider a standard
%EPR set up, as an analogous situation involving only two degrees
%of freedom corresponding to the spin of the two particles rather
%than the infinite set of variables corresponding to the directions
%in the sky. The EPR set-up consist of pair of spin 1/2 particles
%in a common singlet state. The symmetry in  this situation, both
%in the state and in the dynamics,  is the invariance under
%rotation along the axis connecting the two particles.  The
%correlation function for the two observables $\vec S_1. \vec n_1$
%and $\vec S_2. \vec n_2$ is known to be proportional to
%$\cos(\theta)$ where $\theta $ is the angle between $\vec n_1$ and
%$\vec n_2$. The interpretation of this is, that if we consider an
%ensemble of identical systems and
% {\it measure} in each instance the two observables, multiply the
%results, and compute the average over the ensemble the result is
%predicted to be the one given by the two observable correlation
%function. However the prediction in no way indicates a breakdown
%in the symmetry of the situation in the absence of a measurement. This illustrates the aspect of the problem which is not even addressed, in the treatments that concentrate on
%the study of quantum  correlations such as\cite{Correlations}.
}

One may want to interpret step 6 as an effective description of
some sort of measurement process or ``collapse of the wave
function''. This seems to be plausible, as step 6 involves
expectation values of squares of operators, i.e., quantities that
measure the spread of measurement results according to standard
quantum mechanics. Furthermore, a measurement can break the
symmetry of the initial state and produce, in our case,
inhomogeneities. This scenario would be quantum mechanics at its
core: The unitary evolution (the $U$ process in \cite{Penrose}) of
the quantum state with a symmetry preserving Hamiltonian will not
break the initial symmetry of the state, but a measurement of an
observable whose eigenstates are not symmetric, forces the system
(through the collapse, called $R$ process in \cite{Penrose}) into
one of the possible asymmetric states. The onset of the asymmetry
(in standard interpretations of quantum mechanics) occurs only as
a result of the measurement, and then, only when the measured
observable does not commute with the generators of the symmetry.

However this scheme, applied to the situation at hand, immediately
raises many questions: What constitutes the measurement in our
cosmological situation? When did it happen? What is the observable
that was measured?  Perhaps the answer is somehow connected to our
own observation of the sky?  Are we to believe that until 1992
\cite{COBE} the CMB was in fact perfectly homogeneous? Hardly, as
we know the conditions for our own existence depend on the
departures from homogeneity and isotropy in our universe, and thus
our actions could not be their cause.

Moreover, the predictive power of quantum mechanics regarding one
single measurement is rather small (in the EPR case the only absolute
prediction for one experiment concerns to the situation in which the
two measuring axes are perfectly aligned).  How is it, then, that we
have such a predictive power in regards to the single measurement of
our sky?  One could think that this is due to the fact that we are
measuring many regions in the sky and that each should be considered
as a different measurement, but this is not the case: The measurement
of the amplitude in one single mode requires the Fourier transform on
the whole sky. The fact that the actual measurements involve (for
technical reasons) smaller regions than that, is in fact responsible
for uncertainties in the measurement of the quantity of interest.

Step 6 might effectively describe some sort of measurement, but
important questions have to be answered:
\begin{enumerate}
\item What is performing the measurement?
\item Precisely, what is the set of quantum observables
 that is being measured? And what determines them?
\item When is this measurement taking place?
\item Can the above questions be answered in such a way that the
ensuing predictions are in agreement with observations.
\end{enumerate}
%The claim that there exist a prediction
%regarding the primordial density anisotropies and inhomogeneities in the
%universe could not be sustained without addressing these questions.

One further issue that needs discussion is the nature of the
different averages, fluctuations and in general statistical issues
that are present in the problem and that are sometimes treated as
if no differences even of principle did exist. First we have the
quantum mechanical aspect of the problem, reflected for instance,
in the evaluation of expectation values of observables. Here the
statistical nature is reflected in the usual indeterminacy of
future measurements of certain quantities when the state of the
system does not coincide with an eigenstate of the observable.
However let us imagine for a second that we could find an operator
that measured the degree of cosmic inhomogeneity of the system. It
is clear that the vacuum would correspond to an eigenstate of such
operator with zero eigenvalue. If that was the observable that is
measured, the inescapable result would certainly be problematic
for the standard analysis that is supposed to lead to the observed
spectrum of fluctuations. The point is not to discuss whether this
operator exists or not but to note that the statistical aspect
associated with the quantum mechanical picture would emerge only
when the particular observable associated with a measurement is
selected. Next we have the statistics associated with the
classical ensemble represented by the stochastic field and the
corresponding inhomogeneities characterizing the corresponding
ensemble of universes. And finally we have the statistical
description of the inhomogeneities within our own universe, which
can be thought of as an arbitrary generic realization within the
ensemble of universes mentioned before. Here the point is to
distinguish in principle, between the statistics in the ensemble
of universes, and the statistics within one (our own).  We are
used from our experience with statistical mechanics to identify
averages over ensembles with physical expectations, however we
must recall that these identifications rely on two other
identifications: The identification of the ensemble averages with
the long time averages (a fact that relies on the validity of the
ergodic assumptions), and the identification of long time
averages, with the results of physical measurements, a fact that
relies on equilibrium considerations. Needless to say, that none
of these conditions are present in the problem at hand. Some of
these issues have motivated the considerations in
\cite{Correlations}.

The claim that there exist a prediction regarding the primordial
density anisotropies and inhomogeneities in the universe could not
be sustained without addressing these questions.

%-----------------------------------------------------------------------
\section{A look at  proposed ideas on the transition to classicality}
\label{sec_other}
%-----------------------------------------------------------------------

In this section we give a quick overview of the most popular ideas
proposed to address these issues, and attempt to exhibit as
clearly as possible their incompleteness, by signaling the point
at which a ``missing element" makes a disguised appearance, or at
least indicating the place where it should have entered in order
to justify the subsequent interpretation given to the computed
quantities.

%---------------------------------
\subsection*{Standard Decoherence}
%---------------------------------
Decoherence in its standard interpretation describes the fact that
when considering  a quantum system with a very large number of
degrees of freedom, most of which are ignored (by considering them
as ``the environment''), the density matrix for the subset of the
remaining, interesting ``observables'', evolves under certain
circumstances, and  after suitable time averaging, towards a
diagonal matrix. This is sometimes said to represent the emergence
of the classical behavior of the interesting observables. There
has been certainly a large amount of work on this field, most of
it devoted to evaluating the time evolution of the previously
mentioned density matrix in various types of situations, or more
specifically to studying the decoherence functional. There is
however much less work devoted to interpretational issues and it
is fair to say that decoherence has not solved the measurement
problem in quantum mechanics \cite{Adler}. In fact, the diagonal
character of the resulting density matrix point to the
disappearing of certain quantum aspect of the problem. That by
itself is certainly not enough to claim that the situation has
become completely classical: A non-interfering set of simultaneous
coexisting alternatives is not something that can be thought of as
belonging to the realm of classical physics. This sort of
criticism regarding such interpretations of decoherence have of
course been made before, for instance see section, 2.3.4 of
\cite{Nonlocality}. Moreover the diagonal nature of the density
matrix would disappear if we write it using a different basis for
the Hilbert space of the quantum system, clearly indicating that
even this aspect of the so called classicality has a limited
validity.

Thus, the main issues that confront the decoherence approach to the
measurement problem in quantum mechanics are the selection of the
basis, and the fact that after one has a diagonal density matrix, one
is in general, still a big step removed from an interaction specific
eigenstate for the desired observable. In the standard case one can
progress beyond this point by invoking the measurement apparatus to
help select the basis, and by using the ensemble interpretation to
deal with the density matrix. These two aspects are clearly absent in
the cosmological situation we are considering\footnote{As we already
mentioned, taking the position that the measurement is in fact our
own series of studies of the CMB, leads to a closed circle of causes
and effects, with an explanatory power that is highly doubtful to say
the least.}. In fact the need for going beyond standard quantum
mechanics in this context has been noted before \cite{Hartle-1997,
Hartle-2005} { (and we will discuss this line of thought further at
the end of this subsection)}. Furthermore, if one wants to claim that
decoherence solves the problem in the cosmological setting at hand,
one must find a preferential basis selected by a physical mechanism,
and a criterion for the separation of the degrees of freedom into the
``interesting set" and ``the environment" dictated by the physical
problem at hand. { In some schemes one might be able to unify these
two issues into one by arguing that the environment determines the
relevant basis to be that in which the system apparatus environment
interaction is diagonal \cite{Halliwel-89}. In any case, the result
would then naturally have the imprint of these two inputs -- perhaps
unified into one -- but so far no universally accepted prescription
for such choices exits, nor can such selection be fully justified, in
the standard descriptions of the origin of the cosmic density
fluctuations.  For instance, one might want to argue that one needs
to trace over the very large wavelength modes, because they are
unobservable, as is done for instance in \cite{Halliwel-89}, but as
the part of the universe that is observable by a co-moving observer
is certainly dependent on the cosmological  epoch, the tracing would
depend on the cosmic time of the observer, but as such it can not be
argued, unless we do away with  causality, to play a role in the
onset of inhomogeneity and anisotropy in the early universe. Quite
generally, we can not allow our own  characteristics and limitations
to  come in  into the argument,  if our
 hope to explain the asymmetries that give rise to the  conditions that make
 our existence possible.}

Actually, in some of the treatments invoking decoherence one finds
at some point of the discussion, an appeal to a ``a specific
realization" (\cite{Starobinski1} sec. 3, \cite{Parentani} sec. V)
of the stochastic variables, in a clear allusion to something that
can be called the ``collapse" from the statistical description of
the universe, into one of the elements of the statistical
ensemble.\footnote{Note that other schemes are based to some
degree on this sort of specific realizations, without calling them
by this name -- see for instance \cite{Hu}} This is the point, of
course, when we transit from a homogeneous and isotropic
description to an inhomogeneous and anisotropic one, presumably
corresponding to our universe. However nothing is said about when
and how the transition occurs \textit{in our universe}.

{ In fact what seems to be an insurmountable issue in this scheme is
the following: Even if we go from a perfectly symmetric state (the
symmetry being homogeneity and isotropy), to a density matrix for a
subset of the physical degrees of freedom, which is expressed in a
basis in which the non-diagonal elements are negligibly small,
through a  justified  selection of the degrees of freedom that should
be considered environment, one could not
  end up with an
asymmetric mixed state, as there is nothing in the physical laws or
the initial situation that could lead to such breakdown of the
symmetry}. Thus the mixed state described by the density matrix is
still perfectly symmetric.The density matrix is then given a
statistical interpretation, by which we stop regarding the matrix as
representing the state of our universe, and instead view it as
representing a statistical ensemble of universes, among which we find
our own.  The ensemble as a whole is clearly symmetric, but leaves
room for each of the elements in the ensemble to be asymmetric.  This
is how we end up with an inhomogeneous and anisotropic universe.
However, it is clear that this does not address at all any of the
characteristics of transition of our universe from an initial
symmetric state, to such a resulting asymmetric  state. { This
problem affects all the schemes based on decoherence, including the
detailed treatments in \cite{More-Decoherence}, and if we seek a
realistic understanding of the origin of inhomogeneity and anisotropy
in our universe, this approach would clearly be missing something}.

{ After this discussion of the difficulty of applying the standard
picture of decoherence in the cosmological context, we would like to
mention the ideas of Hartle and collaborators. We have placed
them here even though they call for a generalization of Quantum
Mechanics. In fact, we view his arguments for the need of a
generalization of Quantum Mechanics to be applicable not only to
quantum gravity, but also to cosmology
(\cite{Hartle-2005,Hartle-1993}, and references within) as further
indications for the need to go beyond standard physics. The
particular generalization of Quantum Mechanics that has been
suggested involves the assignment of a decoherence functional to
allowed sets of coarse grained histories. One such set of coarse
grained histories decoheres when there is no quantum mechanical
interference among the different alternatives. In that situation we
are able to assign definite probabilities to the particular
alternatives. In the case at hand this would have to be applied to
the histories starting during the inflation period in early universe
to the formation of large structures and eventually to ourselves.

We should however mention some problematic aspects of this proposal:
To start with, the division of the global set of all fine grained
histories into sets of coarse grained histories might be done in
various ways, leading to different decoherent sets \cite{kent}.
Second, the coarse graining, and corresponding decoherence, arises
only when we ignore certain degrees of freedom, and  the
justification for doing so, relies on what  we, as humans living in
this particular era, could in principle observe and what we can not.
This is fine, except in the case that what we seek to understand is
the emergence of the conditions that lead to the possibility of our
own coming into existence.  It is clear that in that case  we can not
call upon some of our own characteristics as a part of the
explanation.}

%--------------------------------------------
\subsection*{Decoherence without Decoherence}
%--------------------------------------------
This can be viewed as a particular realization of the ideas of the
previous case. Its main appeal is that one needs to consider only
the scalar field and no other physical system is required to play
the role of ``environment". The point is that certain modes of the
scalar field vanish asymptotically in time as a result of the
inflationary dynamics of the universe. These modes are then deemed
unobservable, and one is then invited to treat them as such,
taking the appropriate trace and obtaining a mixed state from what
was initially a pure quantum state of the field (ex.
\cite{Starobinski1,Kiefer1}). This approach not only suffers from
the same drawbacks that affect the decoherence approach in
general, but also from what seems as a further interpretational
excess: To have expectation values or higher moments of a mode
converge to the zero value is not the same as making it
unobservable. Zero is just as good a value for a scalar field as
any other one.
%
%   A state with zero angular momentum of an hydrogen
%  atom is no less a state as the others.  In fact for every physical
%  mode we have the standard commutation relations between the
%  corresponding operator and that of its conjugate momentum, thus if
%  the mode is very small in some regime, the corresponding momentum
%  must be appropriately large. Furthermore it is at all unclear how
%  one can one
%  declare that a quantity is or not observable without
%  specifying what is in principle the mechanism of observation that
%  are in principle available.
%
% For instance one can define  a new field $  D= \phi/a^n$ and decide that it
% is unobservable at late times because all its eigenvalues vanish
% asymptotically, however
% the point is that the coupling of $D$ to gravitation is
% enhanced by factors of $a^n$.
%
Moreover it is clear that even if one would agree on the non
observability of certain degrees of freedom, that does not imply
necessarily that the system as a whole needs now to be treated as
a mixed state. There could be other reasons that fix the state of
those degrees of freedom.  A proton can, in the appropriate
circumstances, be described as a pure quantum state, despite the
fact that in these same conditions its microscopic constituents --
quarks and gluons -- might be, in practice, unobservable.

One novel aspect that is sometimes invoked in these treatments is
the squeezing of the vacuum \cite{Starobinski1,Kiefer1}.  That is,
one notes that during the inflationary stage the initial vacuum
state for each mode, evolves towards a squeezed state. We recall
that a squeezed state is a state that has the minimal uncertainty
but not in the standard position and momentum variables but a new
pair of ``rotated" canonical variables.  Thus, in our case, one
has an uncertainty on the value of field and conjugate momentum
which is much larger than the minimum uncertainty provided by
quantum mechanics. Thus one concludes this minimal uncertainty
(indicated by Q.M.) is negligible compared with the actual
uncertainty and thus that one can neglect quantum mechanics.  This
is taken as an indication that there has been a transition from
the quantum to the classical behavior. This conclusion does not
seem to be sufficiently justified, as can be seen by noting that
in many concrete situations, such as those studied in Quantum
Optics, one deals precisely with states that have uncertainties in
conjugate quantities that are larger than the minimal ones
provided by Q.M. and nevertheless there are situations -- for
instance when one is interested in experiments sensitive to
quantities other than the ones in which the system is seen as
squeezed -- that one must take into account that one is dealing
with a system that must be treated quantum-mechanically.
\footnote{The problem can be seen more clearly if we look at the
following example. Take an electron in a minimal wave packet
localized at the origin such that the uncertainty  in the position
$\Delta X = \alpha $ and that in its momentum is $\Delta P =
\hbar/2 \alpha $. Now construct a superposition of such state with
an identical state after a translation by a large distance D.  The
product of the uncertainties is now roughly $ D\hbar/2 \alpha $
which can be made as big as desired by taking $D$ sufficiently
large. Clearly the system is nevertheless far from being
classical.}

We conclude that it is clearly insufficient to have the product of
the uncertainties being much larger than $\hbar$ as criterion for
classicality. One could consider instead as criterion for
classicality the requirement that the volume of phase space
occupied by the system, as measured for instance by the support of
the Wigner functional, be much larger than $\hbar$. This condition
is however not satisfied by a squeezed state. For a critical
discussion of these issues in the context of inflation see also
\cite{Decoherence}\footnote{It is illustrative to mention, in the
context of this example, that one might add to the electrons their
spin degree of freedom and consider the state where the electron
at the origin has its spin up, superposed with the displaced state
where the spin is down. Now consider the density matrix obtained
by ignoring the spin degree of freedom, (and thus tracing over
it). In that case we would have an essentially diagonal density
matrix in the position basis, but one could not argue that one has
a classical situation.}.

And of course we have the other issues that have been pointed out
regarding the general scheme of decoherence as describing the
transition from a quantum, homogeneous, and isotropic state of affairs
to a classical (or quasi-classical), inhomogeneous and anisotropic
situation.

%-------------------------------------
\subsection*{Alternative to Inflation}
%-------------------------------------

In a recent work Wald and Hollands \cite {Inflation-Wald} have
shown that is possible to recover identical predictions as those
afforded by inflation, regarding the spectrum of the seeds for
cosmic structure formation, without requiring the universe to have
gone through an inflationary stage. They start by considering
again the dynamics of a scalar field in a background cosmological
setting. In their model, the inhomogeneities in the universe
originate while the universe is dominated by radiation or dust so
that the scale factor grows as $a(t) =c t^{\alpha}$ with cosmic
time $t$, where $c$ is a constant and $\alpha <1$. Their model
just requires that all modes would have been in their ground state
(as in the standard picture), and that the fluctuations are
``born" in the ground state at an appropriate time which is early
enough so that their physical length is very small compared to the
Hubble radius, and then they ``freeze out" when these two lengths
become equal. The presentation of their model actually allows one
to see clearly the need for some process that would be responsible
for the so called ``birth" of the fluctuations, which can be seen
to play a role similar to that of quantum mechanical measurement.
In fact the so called ``birth of the mode" is the step whereby a
quantum mechanical uncertainty is replaced, in their treatment, by
an actual classical fluctuation of the energy density, and the
point at which a particular mode that has been contributing in an
absolutely homogeneous and isotropic way to the universe density,
becomes a source of the inhomogeneities that presumably are
responsible for the structure in the matter distribution of the
early universe.  The issues are then: What physical process is
responsible for these ``births", or transmutation of the
fluctuations? And how is it that such process selects the
particularly appropriate time of such occurrence for each
particular mode?

%-----------------------------------------------
\subsection*{The `many universes' perspective}
%-----------------------------------------------

One view that is apparently very widespread among the community
working in inflation (but much less represented in the
corresponding literature), is what can be called the Many
Universes perspective. According to this point of view, our
universe is one among a large number of universes that constitute
an ensemble, and it is this ensemble what is in fact characterized
by the quantum state that is homogeneous and isotropic. The
standard treatment is then interpreted as reflecting the most
likely aspect our universe can be expected to have, when
considered statistically within such an ensemble. In such an
interpretation there appear to be no open issues, no need for a
collapse, and certainly not new physics.

Note that this would have to be quite different from the Everett,
or Many Worlds, interpretation of Quantum Mechanics. In the
latter, reality is made of a connected weave of ever splitting
worlds, each one realizing one of the alternatives that is opened
by what we would call a quantum mechanical measurement. It is thus
clear that the points of splitting, the basis in which the
splitting occurs, and the physical entity associated with the
triggering of the splitting are in one to one correspondence with
the aspects we have mentioned before which are associated with the
collapse of the Copenhagen interpretation (time of measurement,
basis, and measuring device). In the Many Universes paradigm there
is no splitting, as that would have amounted effectively to a
collapse and would then be subjected to the equivalent set of
questions: When did the splitting occur? What caused it? And how
was the basis in which it happened, selected? In particular, it
should be emphasized that from such  "Many Universes" point of
view, our universe was never homogeneous and isotropic. Let us
however examine in more detail what such view entails.

In the most direct interpretation of such posture we would have a
collection of universes, each one of which has a definite and
concrete realization set of the asymmetries (the symmetry being of
course homogeneity $+$ isotropy). However, when this is being said,
we clearly imply that each one of these universes would be
susceptible to have a description, in particular, a description of
its asymmetries. That description must be either classical or quantum
mechanical.

If we take the first option, we would be taking the position that
quantum mechanics is a theory applicable to ensembles of systems.
In this view, we must confront two aspects, a) each one of the elements of the ensemble can indeed
be described in a classical language,  and b) the quantum mechanical
description contains information only about the statistics that
results from repeated experimentation applied to the collection of
elements constituting the ensemble. Aspect  b) of this
posture, i.e. the applicability of QM only to ensembles, is in
itself problematic given the absolute nature of the prediction
regarding a single system that can be made in certain special
circumstances (like when dealing with eigenstates of the operators
to be measured or when there are superselection rules totally
forbidding certain processes), for more on this issues see
\cite{Bell}. But even more disquieting is  aspect  a) of the
posture, namely the one where one advocates that each member of
the ensemble has a classical description. This position would
revert one to a sort of ``hidden variable" advocacy, which would
say for instance that in an EPR set-up each electron has a well
defined value of its  spin components even before the measurement
is done, a position which is known to be untenable in light of the
Bell Inequalities  for correlations (See discussion by in \cite{
Mermin}  and their experimental corroboration \cite{Aspect}).

If we take the second option, we would be saying that our universe
is in a quantum state, which is not symmetric, but that, when
superposed with the other asymmetric quantum states that describe
a certain ensemble of universes, leads to an homogeneous and
isotropic quantum state; the one corresponding to the inflationary
field vacuum for the initial stages of inflation. This  seems fine
at first sight, however, it is far from what Quantum Mechanics
prescribes: One does not superpose a state describing one system,
with the state describing a second system to obtain the collective
description of the two systems. A further  problem appears when we
want to take the argument in the opposite direction. Namely,
taking as the initial assumption that the state that collectively
describes the collection of universes is the one corresponding to
the vacuum state of the field and is thus homogeneous and
isotropic, one must end up with a decomposition of that state in a
particular basis (of non-symmetric states) of the Hilbert space in
order to associate to each universe a corresponding quantum state.
Now we face two issues. First, the election of such basis, is not
given ``a priori" by the formalism. One can certainly think of
ways to make the choice, and such choice would correspond to the
selection of the quantity that is ``measured" in ordinary Quantum
Mechanics. This aspect is missing in the standard descriptions. In
fact it would correspond to whatever selects the basis in which
the collapse takes place in our approach. The supra-temporal
selection of the basis (supra-temporal in the sense that, from
this point of view, each universe is from the onset in a specific
non-symmetric quantum state) has no counterpart in ordinary
Quantum Mechanics. For instance when considering an EPR correlated
pair of electrons one could not take the position that there is a
multiplicity of universes and that in ours each electron is in a
specific state of its spin at all times along its path. Similarly,
when considering an electron in its ground state in a hydrogen
atom one would not argue that in our world the electron is in an
eigenstate of the position and that the ground state corresponds
in fact to a description of a corresponding ensemble of universes
with fully localized electrons. This aspect is therefore novel to
the situation at hand, and as such would qualify also as New
Physics.

The second problematic aspect is the following: In taking this
point of view, our universe would be considered as always having
been in a state that is anisotropic and inhomogeneous. It would
have never corresponded to the vacuum state of the scalar field.
Then, one might find uncomfortable the idea that placing our
universe within an ensemble together with a large set of
unobserved universes, one might be able to make predictions about
our particular one. In other words, if we do not assert that our
universe started in the scalar field's vacuum, how could we end up
with prediction for its anisotropy that takes such  a state as the
starting point? Some readers might argue that this is an
epistemological complaint, and dismiss it, while others would
sympathize with the uneasiness. In any event it is worth noting
that this aspect is there.

The final option that seems to be open within this general point
of view asserts that our universe was indeed, together with all
the elements in the ensemble in the start corresponding to the
homogeneous and isotropic vacuum state of the scalar field. That
all these possibilities thus evolved  ``simultaneously" and that
in each one of them (or at least in a great many of them)
structures such as galaxies and stars formed out of the initial
asymmetries, and that observers like ourselves then evolved in
some of the solar systems, and that these observers thus carried
out the observations that selected the basis in which a standard
type of measurement-induced collapse occurred. The problem of
course is that only in very special basis would the galaxies stars
and observers themselves exist to carry out the observations that
effect the measurement-like collapse and concomitant selection of
the basis.

Finally, and as a cautionary note against the hope of finding a
coherent and orthodox description of the situation, one should
keep in mind the fact that the Copenhagen interpretation is
inviable without outside observers. We can not be part of the
system, and our existence can not, thus, be explained in such
scheme. If the purpose of cosmology is to give a picture of the
evolution of our universe that explains the way its present form
and content -- including ourselves -- is arrived at, that
particular point of view will be always lacking.

%---------------------------------------
\section{The missing element}
%and some proposals}
\label{sec_element}
%---------------------------------------
As we have seen, all the scenarios that have been considered are
incomplete and unsatisfactory in some way or another. A close
inspection actually reveals that they are all missing some
element: The process whereby a perfectly homogeneous and isotropic
state (for the universe is homogeneous and isotropic and so is the
vacuum quantum state that one assumes for the scalar field),
transforms into an inhomogeneous and anisotropic state which is
what is described by the density fluctuations. There is clearly no
deterministic mechanism that can achieve this without the
introduction of some external source of asymmetry. Barring such
source, we need to recur to quantum mechanics. However even when
doing so, one is only able to provide for what is required as part
of the so called R process (measurement, collapse, etc.) but not
during the U process (unitary evolution through a
Schr{\"{o}}dinger type equation). Thus if a measurement-like
process is absent there can be no transition. The problem is then
the absence of a sensible candidate for such process. This is
because in the early universe which is homogeneous and isotropic
one can not select a canonical quantity that is the one that would
be measured, and much less the measurement agent or mechanism.
After all, that would require an effective division of the
universe into system and a measuring device, and it is clear that
physics, in this case, does not indicate such division along any
lines we know off.

The main points of this article are, first to indicate that such
aspect is missing and to argue that it requires new physics, and
second to show that by making rather modest suppositions we would
be in a position, thanks to recent advances in observational
status of cosmology, to actually say something non trivial about
the characteristics that this new physical process must possess. A
further speculative discussion about the possible nature of this
new physics will be included.

\subsection*{The standard view, supplemented by the physical collapse
  hypothesis.}

Our approach here is to explore the necessary ingredients to make
the standard picture interpretationally accurate, with a minimal
set of assumptions, and to carefully identify where they occur. We
will assume that there is indeed a \textit{new} physical mechanism
that is responsible for the transformation of the ground state
into a state that contains the fluctuations that are responsible
for the departure from homogeneity and anisotropy. We will call
this process the {\em collapse}. Apart from this the treatment
will be carried out following the standard rules of unitary
quantum mechanical evolution. We will attempt to extract the
conditions that will enable us to recover the standard predictions
of the inflationary model for the appearance of the seeds for
cosmic structure, while exposing the points where we need to
depart from that treatment, in the sense that the uncomfortable
points will not be hidden by the formalism.

The idea is then to follow the standard picture up to the point 6
in Section \ref{sec_standard}, and supplement it by the {\em
collapse hypothesis}. As we mentioned before, we will neglect the
hydrodynamic evolution corresponding to point 7, and assume that
we have direct access to the unmodified spectrum. That is, we will
discuss the conditions under which we could understand the
observation of a scale free Harrison Zeldovich spectrum, directly
in the CMB.

The scheme we have in mind is thus:

1) We split scalar field and metric into ``background'' parts and
perturbations and specify the background.

2) We treat the scalar field perturbation as a quantum field evolving
in the classical space-time background according to the standard
unitary evolution given by its dynamics, except at moments
when gravity (or something else) triggers a collapse of the quantum
state of the field.

3) The collapse itself will be described in a purely
phenomenological manner, without reference to any particular mechanism.

4) We couple the metric perturbations to expectation values of the
scalar field according to semiclassical Einstein equations
\begin{equation}
\label{eq_semi}
G_{ab}=8 \pi G \expec{T_{ab}}
\end{equation}
As implied by 2), we do however neglect the back-reaction of the metric
perturbations on the scalar field evolution because they are
suppressed, as we will show.

We must note that in the moments where collapse does occur, the
semiclassical Einstein equations (\ref{eq_semi}) are
violated.\footnote{We thank Prof. A. Ashtekar for pointing this
out.} The view we take here is that in these moments a more
fundamental description, possibly a theory of quantum gravity,
would provide the correct equations for the collapse. Thus this is
certainly a very schematic picture that must be further studied,
analyzed, and specified, if we want to construct a complete model.
At this time we are interested only in finding out the basic
aspects that the model should have in order to actually account
for the observational facts. On the other hand we should point out
that the scheme is certainly inspired by the ideas of Penrose
\cite{Penrose} regarding the role of quantum gravity in the
collapse of the wave function in general circumstances, and that
ideas of this sort of mechanism have been proposed quite
independently from any quantum gravity consideration, in the
context of the interpretational problems of Quantum Mechanics
\cite{GRW}.

We proceed now to give a more detailed analysis. In particular,
and in order to be able to discuss more clearly the ideas related
to the collapse proposal we will decouple the treatment of the
gravitational degrees of freedom from those of the true scalar
field which will be quantized in a standard way.

%-------------------------------------------------------------------
\section{Linearized Einstein's equations and the evolution of small
  fluctuations}
\label{sec_linear}
%------------------------------------------------------------------

We will consider here the metric and field perturbations taken as
classical test fields evolving in the background geometry. As the
metric (\ref{RW}) is conformally Minkowski, it is convenient (and
customary) to go over to a new time coordinate $\eta$ (`conformal
time') that makes this explicit.  To make things more definite we
set $t=0$ and $\eta=0$ to mark the time at which inflation ends.
Thus we have $\eta \equiv \int_0^t a^{-1} (t') dt'$. To finalize
fixing our conventions we set the scale factor to be 1 at the
present time $t_0$. The background metric is then

\begin{equation}
ds^2=a^2(\eta)[-d\eta^2+(d \vec x)^2].
\end{equation}
During inflation the scale factor is given by
\begin{equation}
a(t)=d \exp(H_{\rm I}\ t),
\end{equation}
where $H_{\rm I}$ is the Hubble constant during
inflation.
Thus for $t<0$ we have
\begin{equation}
\eta(t)=\frac{1}{H_{\rm I}\ d}(1- e^{-H_{\rm I} t} ),\qquad
a(\eta)=-\frac{1}{H_{\rm I} (\eta-\eta_0)}.
\label{expansion}
\end{equation}
where $\eta_0\equiv \frac{1}{ H_{\rm I}d}$.  Let us reparametrize
the evolution fixing $\eta_0=0$. The point is then that inflation
would end at some $\eta_{IE}<0$ and afterwards the universe will
proceed to a standard cosmological expansion until at $\eta=0$ $a=1$.
We will ignore this part of the universe evolution in the rest of the
paper.

It is customary to decompose the metric fluctuations in terms of
its scalar, vector, and tensor components. In the case of our
Einstein-inflaton system only scalar (generated by density
perturbations) and tensor perturbation (gravitational waves) are
relevant. Gravitational waves will be ignored for simplicity and
thus, the perturbed metric (in the conformal gauge) can be simply
written as
\begin{equation}
ds^2=a(\eta)^2\left[-(1+ 2 \Phi) d\eta^2 + (1- 2
\Psi)\delta_{ij} dx^idx^j\right],
\end{equation}
where $\Phi$ and $\Psi$ are scalar fields, the former is referred
to as the Newtonian potential.

Let us first write down the components of the Einstein tensor
($G_{ab}=R_{ab}-\frac{1}{2}g_{ab}R$) up to first order in the
perturbations: \ba
&& \nonumber  G^{(0)}_{00}=3\frac{\dot a^2}{a^2} \\
\nonumber && G^{(0)}_{ii}=\frac{\dot a^2}{a^2}-2\frac{\ddot a}{a} \\
\nonumber &&
G^{(1)}_{00}=2 \nabla^2 \Psi - 6\frac{\dot a}{ a} \dot \Psi \\
\nonumber && G^{(1)}_{0i}=2 \partial_{i} \dot \Psi+ 2 \frac{\dot
  a}{ a} \partial_{i} \Phi\\
\nonumber && G^{(1)}_{ii}=( \nabla^2 -\partial_i \partial_i
)(\Phi-\Psi) + 2 \ddot \Psi + 2\left( 2 \frac{\ddot a}{a}-\frac{\dot
    a^2}{a^2}\right)(\Psi+\Phi)+2 \frac{\dot a}{a} (\dot \Psi+ \dot
\Phi)\\ && G^{(1)}_{ij}=\partial_i \partial_j (\Psi-\Phi) \ \ \ {\rm
  for} \ \ \ i \not= j \label{eeee}\ea The components of the energy momentum
tensor are as follows \ba \label{tttt}
&& \nonumber T^{(0)}_{00}=\frac{1}{2} (\dot \phi^2_0) +a^2 V[\phi_0]\\
\nonumber && T^{(0)}_{ii}=\frac{1}{2} (\dot \phi^2_0) -a^2
V[\phi_0]\\
\nonumber && T^{(1)}_{00}=\dot \phi_0 \delta\dot\phi +
2 a^2 \Phi V[\phi_0]+ a^2\partial_{\phi}V[\phi] \delta\phi\\
\nonumber &&
T^{(1)}_{0i}=\dot \phi_0 \partial_i \delta\phi\\
\nonumber && T^{(1)}_{ii}=-\Phi \dot \phi^2_0 +\dot \phi_0
\delta\dot\phi- \Psi (\dot \phi^2_0-a^2 V[\phi_0])-\frac{1}{2} a^2
\partial_{\phi}V[\phi] \delta\phi\\
&& T^{(1)}_{ij}=0\ \ \ {\rm for} \ \ \ i \not= j \ea Finally the
scalar field equation yields, to zeroth order:
\begin{equation}
\ddot\phi_0 + 2 \frac{\dot a}{ a}\dot\phi_0 +
a^2\partial_{\phi}V[\phi] =0, \label{Scalar0}
\end{equation}
and to first order:
\begin{equation}\ddot\delta\phi + 2 \frac{\dot a}{ a}\dot\delta\phi  -
\nabla^2 \delta\phi + a^2\partial^2_{\phi,\phi}V[\phi] \delta\phi
-( \dot \Phi + 3 \dot  \Psi ) \dot\phi_0  -2 \Psi  (\ddot\phi_0 +2
\frac{\dot a}{ a}\dot\phi_0)=0.
\end{equation}
Now let us reduce the number of equations by solving some of them.
The only non-trivial among Einstein's equations, to zeroth order
is $G^{(0)}_{00}=8\pi G T^{(0)}_{00}$ which leads to Friedman's
equation
\begin{equation}
3\frac{\dot a^2}{a^2}=4\pi G  (\dot \phi^2_0+ 2 a^2 V[\phi_0]).
\end{equation}
In the linear order let us start from $G^{(1)}_{ij}=8\pi G
T^{(1)}_{ij}$ which implies
%the metric perturbation potential to be
%equal, namely
$ \Psi=\Phi$.
%\begin{equation}
%\label{popo} \Psi=\Phi
%\end{equation}
Using the previous result, the vector constraint equations
$G^{(1)}_{0i}=8\pi G T^{(1)}_{0i}$ imply
\begin{equation}
\partial_{i} (\dot \Psi+ \frac{\dot a}{ a} \Psi -4\pi G  \dot \phi_0
\delta\phi)=0
\end{equation} which reduces to\footnote{
This, and some of the following results, follow strictly only for
appropriate boundary conditions. In later parts of the paper we
will be working with Fourier decomposition of the various
equations and the vanishing of the corresponding equations for
each of the Fourier mode (except the $k=0$ mode) do follow
independently of any considerations involving boundary
conditions.}
\begin{equation}
\dot \Psi = -\frac{\dot a}{ a} \Psi + 4\pi G  \dot \phi_0
\delta\phi. \label{pd}
\end{equation}
The scalar constraint equation $G^{(1)}_{00}=8\pi G T^{(1)}_{00}$
becomes
\begin{equation}
2 \nabla^2 \Psi
- 6\frac{\dot a}{ a} \dot \Psi = 8\pi G (\dot \phi_0
\delta\dot\phi + 2 a^2 \Psi V[\phi_0]+ a^2
\partial_{\phi}V[\phi] \delta\phi).
\label{1}
\end{equation}
Now, using Einstein's equations to express the potential
\begin{equation}  2 a^2V[\phi_0]= T_{00}^{(0)}-T_{ii}^{(0)} =(8\pi G)^{-1}
(G_{00}^{(0)}-G_{ii}^{(0)})=(4\pi G)^{-1}(\frac{\dot a^2}{a^2} +
\frac{\ddot a}{a}),
\end{equation}
and substituting this result and the value of $\dot \Psi$ from
(\ref{pd}) in equation (\ref{1}), we obtain
\begin{equation}
\nabla^2 \Psi = 4\pi G ( 3\frac{\dot a}{ a}  \dot \phi_0 \delta\phi
+\dot \phi_0 \delta\dot\phi +  a^2
\partial_{\phi}V[\phi] \delta\phi) + (\frac{\ddot a}{a} -2{\frac{ \dot
a}{a}}^2 )\Psi.
\end{equation}
We write this equation as
\begin{equation}
\label{25} \nabla^2 \Psi  -\mu \Psi= 4\pi G ( u \delta\phi +\dot
\phi_0 \delta\dot\phi )
\end{equation}
where $\mu\equiv ( 2{\frac{ \dot a}{a}}^2-\frac{\ddot a}{a} )$,
and $ u\equiv 3\frac{\dot a}{ a} \dot \phi_0 + a^2
\partial_{\phi}V[\phi]$, which upon use of the expression for
$\partial_{\phi}V[\phi]$ from (\ref{Scalar0}), gives $ u= \frac{\dot
  a}{ a} \dot \phi_0 - \ddot\phi_0 $.

Finally using the expressions for the scale factor during
inflation we find $\mu=0$, while the slow-rolling approximation
$\frac{\partial^2
  \phi}{\partial t ^2} =0$ corresponds in these coordinates to the
condition $u=0$. Thus the last equation becomes
\begin{equation}
\nabla^2 \Psi  = 4\pi G \dot \phi_0 \delta\dot\phi. \label{main2}
\end{equation}
We will work during the rest of the paper with the slightly more
general case that corresponds to maintaining the $\mu$ term to
account for possible slight departures of the exponential
expansion in physical time that is reflected in expression
(\ref{expansion}), but will drop the $u$ term indicating that we
are keeping the slow roll regime approximation as important. The
reason for this will become clear when we will carry out the
comparison with the observations.

Now, let us take a look at the evolution equation of the scalar
field fluctuations; if we use (\ref{pd}) in the equation for the
scalar field perturbation we obtain
\begin{equation}
\ddot\delta\phi + 2 \frac{\dot a}{ a}\dot\delta\phi  -
\nabla^2 \delta\phi + a^2\partial^2_{\phi,\phi}V[\phi]\delta \phi
- 16 \pi G  (\dot\phi_0)^2 \delta\phi  -2 \Psi \ddot\phi_0=0
\end{equation}
while using the slow rolling approximation we get
\begin{equation}
\label{eq_fieldeq}
\ddot\delta\phi + 2 \frac{\dot a}{ a}\dot\delta\phi  -
\nabla^2 \delta\phi + a^2\partial^2_{\phi,\phi}V[\phi]\delta \phi
- 16 \pi G  (\dot\phi_0)^2 \delta\phi  -2 \Psi \frac{\dot a}{ a}
\dot\phi_0=0.
\end{equation}
This differs from the evolution equation of the scalar field
perturbations in the background space-time. However note that the
corrections due to the Newtonian potential $\Psi$ are suppressed
by the factor $G$. Thus in our present treatment we will ignore
the complications of maintaining the terms that could be
considered as reflecting the effect on the field of the metric
response to the fluctuations of the field itself.  In fact, from
our point of view the metric would be unchanged until the state of
the scalar field collapses, and only after that would the metric
be changed and could therefore have a back reaction on the
evolution of the field modes.

Our main equations will be then the equation for the scalar field in
the background space-time
\begin{equation}
\ddot\delta\phi + 2 \frac{\dot a}{
a}\dot\delta\phi  - \nabla^2 \delta\phi +
a^2\partial^2_{\phi,\phi}V[\phi]\delta \phi =0
\end{equation}
and equation (\ref{25}), which, upon quantization of the scalar field
perturbation $\delta\phi$  will be  promoted  to a
semi-classical equation to determine $\Psi$ in terms of $<\hat
\delta\phi>$. We will come back to this later.
%-------------------------------------------------------------------
\section{Quantum theory of fluctuations}
\label{sec_quantum}
%-------------------------------------------------------------------

Considering the previous items let us present the standard
treatment of this topic including a brief review of the usual
discussion of the amplification of fluctuations by inflation. We
follow in some of the discussion mainly \cite{branderberger} and
\cite{Liddle}.\footnote{Regarding units we use a convention where
the coordinates will have dimensions of length $(c=1)$, $a(\eta)$
is dimensionless so $H_I$ has dimensions of inverse length. The
scalar field has units of $(Mass/Length)^{1/2}$ and Newton's
constant $G$ has dimensions of $(Length/Mass)$.}

The  starting point is the  assumption is that the dynamics of the
universe, during the inflationary period, is dominated by the so
called inflaton field. The inflaton field is a scalar field
described by the action
\begin{equation}
S[\phi]=\int \left[-\frac{1}{2} \nabla_a \phi \nabla_b \phi g^{ab}
- V[\phi] \right]\sqrt{-g} {\rm d}^4x. \label{act}
\end{equation}
Using the form of the FRW line element the previous action becomes
\begin{equation}
\label{action} S[\phi]=\int  \left[\frac{a^2}{2} (-\phi
\partial_0 \partial_0 \phi + \phi \partial_i \partial_j \phi
\delta^{ij}) - a^4 V[\phi] \right] {\rm d}^4x.
\end{equation}
In order for inflation to take place some hypothesis about the initial
condition for the scalar field and its potential have to be stated.
 %It is customary to separate the scalar degrees of freedom into an
%homogeneous background component $\phi_0$ plus small fluctuation
%$\delta \phi$, namely
%\begin{equation}
%\phi=\phi_0+\delta\phi.
%\end{equation}
One writes  the field as  $\phi=\phi_0+\delta\phi$, where the
background field $\phi_0$ is described in a completely classical
fashion while only the fluctuation $\delta\phi$ is quantized.

In these coordinates, the field equation becomes
\begin{equation}
\delta \ddot \phi-\nabla^2 \delta \phi +2\frac{\dot a}{a}\delta
\dot\phi=0
\end{equation}
where dots denote derivatives with respect to $\eta$ and $\Delta$ is
the Laplacian on Euclidean three space (whose metric is $dr^2+r^2
d\Omega^2$). Note that we have neglected a terms proportional to
$\partial^2_{\phi^2}V[\phi]$ using the slow rolling approximation.

If we expand the fluctuation in its Fourier components the equations
of motion for the mode $\delta \phi_k$ becomes
\begin{equation}
\label{fluctu1} \delta \ddot\phi_{k} + 2\ \frac{\dot a}{a}\ \delta
\dot \phi_{k} -k^2 \delta \phi_{k} = 0.
\end{equation}
It is well known that this equation can be further simplified by going
over from $\delta \phi$ to an auxiliary field $y=a\delta \phi$.  In
the resulting equation for $y$, there is no term with a first
derivative of the field anymore,
\begin{equation}
\label{equ1a}
\ddot y-\left(\nabla^2 +\frac{\ddot a}{a}\right)y=0,
\end{equation}
and in the inflationary regime with $a(\eta)=-(H_{\rm
I}\eta)^{-1}$ it can be explicitly solved. Obviously, a
quantization $\widehat{y}$ of $y$, will immediately give a
quantization $\widehat{\delta\phi}=a^{-1}\widehat{y}$ of
$\delta\phi$, so we will now proceed to quantize the auxiliary
field $y$.

In order to avoid infrared problem we introduce a regularization and
consider the field in a box of side $L$ decompose a real
classical field $y$ satisfying (\ref{equ1a}) into plane waves
\begin{equation}
y(\eta,\vec{x})=\frac{1}{L^{3}} \Sigma_{ \vec k} \left(a_k(\eta)
e^{i\vec{k}\cdot\vec{x}}+ \cc{a}_k(\eta)
e^{-i\vec{k}\cdot\vec{x}}\right),
\end{equation}
where the sum is over the wave vectors $\vec k$ satisfying $k_i L=
2\pi n_i$ for $i=1,2,3$ with $n_i$ integers. The coefficients
$a_k(\eta)$ satisfy the equation
\begin{equation}
\label{equ2a}
 \ddot a_k+\left(k^2-\frac{\ddot a}{a}\right)a_k=0.
\end{equation}
We quantize $y$ field imposing standard commutation relations between
the field and its canonical conjugate momentum
$\py=\dot{y}-y\dot{a}/a$ (which is equivalent to imposing these
relations on $\phi$ and its conjugate momentum $\pi=a^2\dot{\phi}$).
Thus we write
\begin{equation}
\y(\eta,\vec{x})=
\frac{1}{L^{3}}\Sigma_{\vec k} \left(\ann_k(\eta)
e^{i\vec{k}\cdot\vec{x}}+\cre_k(\eta)
e^{-i\vec{k}\cdot\vec{x}}\right),\ \ \ {\rm where}\ \ \
\ann_k(\eta)=y_k(\eta)\ann_k,
\end{equation}
$y_k(\eta)$ is a solution of (\ref{equ2a}) and $\ann_k$ is the
usual annihilation operator on the one particle Hilbert space ${\cal H}={\cal
  L}^2(L^3,d^3x)$. Upon choosing the solutions $y_k(\eta)$,
$\widehat{y}$ thus becomes an operator on the Fock space over ${\cal
  H}$.  Similarly the canonical conjugate to $y$ is given by
\begin{equation}
\py(\eta,\vec{x})=\frac{d}{d\eta} {\hat y}(\eta,\vec{x})- \frac{\dot a}{a}\
\y(\eta,\vec{x})
\end{equation}
can be written as;
\begin{equation}
\py(\eta,\vec{x})=
\frac{1}{L^{3}}\Sigma_{\vec k}, \left(\ann_k g_k(\eta)
e^{i\vec{k}\cdot\vec{x}}+\cre_k g_k(\eta)
e^{-i\vec{k}\cdot\vec{x}}\right),
\end{equation}
where $g_k=\dot y_k-\frac{\dot a}{a}y_k$.
%\begin{equation}
%\label{eq_gmode}
%g_k=\dot y_k-\frac{\dot a}{a}y_k.
%\end{equation}
To complete the quantization, we have to specify the classical
solutions $y_k(\eta)$. This choice is not completely free: To insure
that canonical commutation relations between $\y$ and $\py$ give
$[\ann_k, \cre_{k'}] =\hbar L^3\delta_{k,k'}$, they must satisfy
\begin{equation}
y_k(\eta)\cc{g_k}(\eta)-\cc{y_k}(\eta)g_k(\eta) =-i
\end{equation}
for all $k$ at some (and hence any) time $\eta$. The choice of the
$y_k(\eta)$ corresponds to the choice of a vacuum state for the field,
which in the present case, as on any non stationary space-time, is not
unique. However, we must emphasize that, at this point any such
  selection of a vacuum (made through the choice of the $y_k(\eta)$'s
  that we take as positive energy modes), would be a spatially
  homogeneous and isotropic state of the field, as can be seen by
  evaluating directly the action of a translation or rotation
  operators (associated with the hypersurfaces $\eta =$ {\it constant}
  of the background space-time) on the state.  We will use a rather
  natural candidate for such a
state, the so called Bunch-Davies vacuum. It is characterized by
the choice
%
%  vacuum and we will specify and discuss
% our choice below.
% The pair of independent solutions we take is given by
\begin{equation}
y^{(\pm)}_k(\eta)=\frac{1}{\sqrt{2k}}\left(1\pm\frac{i}{\eta
k}\right)\exp(\pm i k\eta),
\label{Sol-y}
\end{equation}
and
\begin{equation}
g^{\pm}_k(\eta)=\pm
i\sqrt{\frac{k}{2}}\exp(\pm i k\eta) \label{Sol-g}
\end{equation}
Note that the form of $y_k^+$ reduces near $\eta=-\infty$ to that
of the standard positive frequency solution in flat space. This
constitutes our choice of the vacuum of the theory.\footnote{Note
that the dimensions implied for $\ann_k$ is $
(Mass)^{1/2}(Length)^2$ which is compatible with the
dimensionalized commutator $[\ann_k, \cre_{k'}] =\hbar
L^3\delta_{k,k'}$.}

It is convenient to rewrite the field and momentum operators in as
\begin{equation}
\y(\eta,\vec{x})=
 \frac{1}{L^{3}}\sum_{\vec k}\ e^{i\vec{k}\cdot\vec{x}} \hat y_k(\eta), \qquad
 \py(\eta,\vec{x}) =
\frac{1}{L^{3}}\sum_{\vec k}\ e^{i\vec{k}\cdot\vec{x}} \hat \pi_k
(\eta)
\end{equation}
where $\hat y_k (\eta) \equiv y_k(\eta) \ann_k +\bar y_k(\eta)\cre_{-k}$
% and similarly
%\begin{equation}
%\py(\eta,\vec{x}) =
%\frac{1}{L^{3}}\sum_{\vec k}\ e^{i\vec{k}\cdot\vec{x}} \hat \pi_k
%(\eta)
%\end{equation}
% where
and  $\hat \pi_k (\eta) \equiv g_k(\eta) \ann_k + \bar g_{k}(\eta)
\cre_{-k}$.

Furthermore we will decompose both $\hat y_k (\eta)$ and $\hat \pi_k
(\eta)$ into their real imaginary parts $\hat y_k (\eta)=\hat y_k^R
(\eta) +i \hat y_k^I (\eta)$ and $\hat \pi_k (\eta) =\hat \pi_k^R
(\eta) +i \hat \pi_k^I (\eta)$ where
\begin{equation}
\hat y_k^R (\eta) =
\frac{1}{\sqrt{2}}\left(
 y_k(\eta) \ann_k^R
 +\bar y_k(\eta) \creR_k\right) ,\qquad  \hat y_k^I (\eta)
=\frac{1}{\sqrt{2}}\left(
 y_k(\eta) \ann_k^I
 +\bar y_k(\eta) \creI_{k}\right)
\end{equation}
\begin{equation}
\hat \pi_k^R (\eta) =\frac{1}{\sqrt{2}}\left( g_k(\eta)
\ann_k^R
 + \bar g_{k}(\eta) \creR_{k} \right),
\qquad \hat \pi_k^I (\eta) =\frac{1}{\sqrt{2}}\left( g_k(\eta)
\ann_k^I
 + \bar g_{k}(\eta) \creI_{k} \right)
\end{equation}
where
\begin{equation}
\ann_k^R\equiv \frac {1}{\sqrt 2} ( \ann_k +\ann_{-k}), \qquad
\ann_k^I\equiv \frac {-i}{\sqrt 2} ( \ann_k -\ann_{-k})
\end{equation}
We note that the operators $\hat y_k^{R, I} (\eta)$ and $\hat
\pi_k^{R, I} (\eta)$ are therefore hermitian operators.  The
commutation relations of the real and imaginary creation and
annihilation operators are however nonstandard:
\begin{equation}
\left[\ann_k^R,  \creR_{k'}\right] = \hbar L^3 (\delta_{k,k'} +\delta_{k,-k'}), \qquad
\left[\ann_k^I,  \creI_{k'}\right] = \hbar L^3
(\delta_{k,k'} -\delta_{k,-k'})
\label{NewCom1}
\end{equation}
%\begin{equation}
%\left[\ann_k^I,  \creI_{k'}\right] = \hbar L^3
%(\delta_{k,k'} -\delta_{k,-k'}) \label{NewCom2}
%\end{equation}
with all other commutators vanishing.  This is known to indicate that
the operators corresponding to $k$ and $-k$ are identical in the real
case (and identical up to a sign in the imaginary case).

%-------------------------------------------------------------------
\section{Evolution of the fluctuations through collapse}
\label{sec_collapse}
%-------------------------------------------------------------------

In this section, we will specify our model of collapse, and do the
necessary computations to follow the field evolution through collapse
to the end of inflation.

{\bf The collapsing modes }\newline To describe the collapse of
the state the scalar field is in, we will use the decomposition of
the field into modes. It is imperative that these modes are
`independent', i.e. that they give a corresponding decomposition
of the field operator into a sum of commuting `mode operators', an
orthogonal decomposition of the one-particle Hilbert-space, and a
direct-product decomposition for the Fock space. Furthermore, we
require that the initial state of the field is not an entangled
state with respect to this decomposition, i.e. it can be written
as a direct product of states for the mode operators. This ensures
that the notion of ``collapse of an individual mode" will make
sense.

Although the above requirements place some restrictions, there are
different possible choices for this decomposition and the
corresponding choice of modes, with possibly observable
consequences. Until a less phenomenological description of the
collapse becomes available, we can only be guided by simplicity,
and the condition that the end result of our calculation be
compatible with astrophysical observations. Throughout the main
text, we will use the modes labeled by the wave vector $k$ and the
superscript R/I in the last section. We will however show that
there is a certain degree of robustness of predictions under
change of the mode decomposition, by repeating the calculations
that will follow below -- with a different choice of decomposition
-- in appendix \ref{app_spherical}.

Let us be more precise: We will assume that the collapse is
somehow analogous to an imprecise measurement of the operators
$\hat y_k^{R, I} (\eta)$ and $\hat \pi_k^{R, I} (\eta)$ which, as
we pointed out, are hermitian operators and thus reasonable
observables.  These field operators contain complete information
about the field.  As we have to follow the evolution of these
modes during inflation, let us collect some formulas for the
evolution of their lowest moments: Let $|\Xi\rangle$ be any state
in the Fock space of $\hat{y}$. Let us introduce the following
quantities:
\begin{equation}
\label{equ3}
 d_k^{R,I} = \expec{\ann_k^{R,I}}_\Xi,\qquad
c_k^{R,I}=\expec{(\ann_k^{R,I})^2}_\Xi,\qquad
 e_k^{R,I} = \expec{\creRI_k\ann_k^{R,I}}_\Xi.
\end{equation}
In terms of these, the expectation values of the modes are expressible
as
\begin{equation}
\expec{\y_k^{R,I}}_\Xi = \sqrt{2}\re(y_k d_k^{R,I})  \qquad,
\expec{\pyRI_k}_\Xi = \sqrt{2}\re(g_k d_k^{R,I})
\end{equation}
while their corresponding dispersions are
\begin{equation}   \fluc{\y_k^{R,I}}_\Xi
= \re(y_k^2 c_k^{R,I})+ (1/2)\betr{y_k}^2(\hbar
L^3+2e_k^{R,I})-2\re(y_k d_k^{R,I})^2
\end{equation}
and
\begin{equation}
\fluc{\pf_k^{R,I}}_\Xi
=\re(g_k^2 c_k^{R,I})+(1/2)\betr{g_k}^2(\hbar L^3
+2e_k^{R,I})-2\re(g_k d_k^{R,I})^2
\end{equation}
For the vacuum state $|0\rangle$ we certainly have $
d_k^{R,I}=c_k^{R,I}=e_k^{R,I}=0$, and thus
\begin{equation}
\expec{\y_k^{R,I}}_0 = 0,  \qquad  \expec{\pyRI_k}_0 =0,
\end{equation}
while their corresponding dispersions are
\begin{equation}\label{momentito}
\fluc{\y_k^{R,I}}_0 =(1/2) \betr{y_k}^2(\hbar L^3), \qquad
\fluc{\pf_k^{R,I}}_0 =(1/2)\betr{g_k}^2(\hbar L^3).
\end{equation}

{\bf The collapse}\newline

Now we will specify the rules according to which collapse happens.
Again, at this point our criteria will be simplicity and naturalness.
Other possibilities do exist, and may lead to different
predictions. To illustrate this point, in appendix
\ref{app_alternative} we will explore an alternative model.

What we have to describe is the state $|\Theta\rangle$ after the
collapse. To keep things simple and general, we will not  consider specifying the
state completely, but only the expectation values
\begin{equation}
\label{newstate}
d^{R,I}_{k, c} = \expec{\ann_k^{R,I}}_\Theta,\qquad
c^{R,I}_{k,c}=\expec{(\ann_k^{R,I})^2}_\Theta,\qquad
e^{R,I}_{k,c} = \expec{\creRI_k \ann_k^{R,I}}_\Theta.
\end{equation}
where the subscript $c$ indicates that we are talking about the
post collapse values, to distinguish them from their pre collapse
values that as we said are zero. We will drop that subscript in
the following.

At this point a few remarks on our statistical treatment are in
order. We view the collapsed state of the field corresponding to
our universe to be a single state $|\Theta\rangle$ and not in any
way an ensemble of states.  This would seem to result in a
difficulty in principle for the attempts to apply statistical
analysis in our situation, and it reflects a previously mentioned
difficulty of principle that arises when dealing with the fact
that we need a quantum treatment but we have just one universe at
our disposal. However it is an issue we must face if we want to
have a clear and realistic understanding of the issues at hand.
The way we address this issue is related to the fortunate
situation that we do not measure directly and separately the modes
with specific values of $\vec k$, but rather an aggregate
contribution of all such modes to the spherical harmonic
decomposition of the temperature fluctuations on the celestial
sphere. In order to proceed we construct an imaginary ensemble of
universes. Thus we have an ensemble of universes characterized by
the after collapse state $|\Theta\rangle_{i}$ where the label $i$
identifies the specific element in the ensemble . Then we will
have an independent random series of numbers $q^{(i)}_{\vec k}$
pertaining to value of physical quantities in the collapsed state
in each element $i$ in the ensemble for every single $\vec k$ (we
will be assuming there are no correlations among the various
harmonic oscillators).  Our universe however, corresponds to a
single element $i_0$ in the ensemble, leading to the choice for
each $\vec k$ of a number $q^{(i_0)}_{\vec k}$ from among those
random sequences.  The point is that the complete sequence that
corresponds to our universe -- i.e. the sequence $q^{(i_0)}_{\vec
k}$ of specific quantities for fixed $i_0$ but for the full set of
$\vec k$ -- will, as a result, also be a random sequence.

In our specific calculation this approach will be taken with
respect to the quantities $c^c_k,d^c_k,e^c_k$ after considering
the issues of relative overall normalization of the random
sequences. In fact in this first treatment we will not concern
ourselves with the quantities $c^c_k,e^c_k$ which are related to
the spread of the wave packet after collapse, as this will be the
subject of future research.

Thus we focus on  specifying $d_k^c$: In the vacuum state, $\y_k$ and
$\py_k$ individually are distributed according to Gaussian
distributions centered at 0 with spread $\fluc{\y_k}_0$ and
$\fluc{\py_k}_0$ respectively. However, since they are mutually
non-commuting, their distributions are certainly not independent.  In
our collapse model, we do not want to distinguish one over the other,
so we will ignore the non-commutativity and make the following
assumption about the (distribution of) state(s) $|\Theta\rangle$ after
collapse:
\begin{equation}
\expec{\y_k^{R,I}(\eta^c_k)}_\Theta=X^{R,I}_{k,
1},\qquad\expec{\pyRI_k(\eta^c_k)}_\Theta=X^{R,I}_{k,2}
\end{equation}
where $X^{R,I}_{k,1},X^{R,I}_{k,2}$ are random variables, distributed
according to a Gaussian distribution centered at zero with spread
$\fluc{\y_k^{R,I}}_0$, $\fluc{\pyRI_k}_0$, respectively. Another way
to express this is
\begin{eqnarray}
\expec{\y_k^{R,I}(\eta^c_k)}_\Theta&=&x^{R,I}_{k,1}
\sqrt{\fluc{\y^{R,I}_k}_0}=x^{R,I}_{k,1}|y_k(\eta^c_k)|\sqrt{\hbar L^3/2},\\
\expec{\pyRI_k(\eta^c_k)}_\Theta&=&x^{R,I}_{k,2}\sqrt{\fluc{\pyRI_k}
_0}=x^{R,I}_{k,2}|g_k(\eta^c_k)|\sqrt{\hbar L^3/2},
\end{eqnarray}
where $x_{k,1},x_{k,2}$ are now distributed according to a Gaussian
distribution centered at zero with spread one.

We now take these equations and solve for $d^{R,I}_k$. Defining the
angles $\alpha,\beta,\gamma$ as
$\alpha^{R,I}_k=\arg(d^{R,I}_k)$,$\beta_k=\arg(y_k )$,
$\gamma_k=\arg(g_k )$, where the last two refer to quantities
evaluated at the collapse time $\eta^c_k$, the above equations can be
written
\begin{equation}
|d^{R,I}_k|
\cos(\alpha^{R,I}_k+\beta^c_k)=\frac{1}{2}x^{R,I}_{k,1}\sqrt{\hbar
L^3},\qquad |d^{R,I}_k|
\cos(\alpha^{R,I}_k+\gamma^c_k)=\frac{1}{2}x_{k,2}\sqrt{\hbar
L^3}. \label{Col2}
\end{equation}
The general solution gives:
\begin{equation}
|d^{R,I}_k| =\sqrt{\hbar L^3}D^{R,I}_k \label{Co3}
\end{equation}
with
\begin{equation}
D^{R,I}_k =(1/2)\frac{(1+z_k^2)^{1/2}}{z_k} ({x^{R,I}_{k,1}}^2
+{x^{R,I}_{k,2}}^2
-2{x^{R,I}_{k,1}}{x^{R,I}_{k,2}}(1+z_k^2)^{-1/2})^{1/2}
\label{Col4}
\end{equation}
where $z_k\equiv k \eta^c_k$, and
\begin{equation}
\cos(\alpha^{R,I}_k +z_k-\pi/2)= x^{R,I}_{k,2}/ (2D^{R,I}_k)
\label{Col5}
\end{equation}
In order to more fully specify the state $|\Theta\rangle$ we would
need also to consider the quantities
\begin{equation}
\label{equ5}
 \delta{(\y^{R,I})}_k = {\fluc{\y^{R,I}_k}_\Theta(\eta^c_k)},\qquad
 \delta{(\pyRI)}_k = {\fluc{\pyRI_k}_\Theta(\eta^c_k)}
\end{equation}
To specify these, and their time evolution, we would have to
specify the remaining six parameters $\re(c^{R,I}_k)$,
$\im(c^{R,I}_k)$, and $e^{R,I}_k$. However, the limited set of
results that are the concern of this paper are independent of such
choice, and we will not further discuss the higher moments of
$|\Theta>$ in what follows.

We need to concentrate on the expectation value of the quantum
operator which appears in our basic formula
\begin{equation}\nabla^2 \Psi -
\mu  \Psi= s \Gamma \label{main3}
\end{equation}
(where we introduced the abbreviation $s=4\pi G \dot \phi_0$) and
the quantity $\Gamma$ as the aspect of the field that acts as a
source of the Newtonian potential.  In a general situation
$\Gamma= \delta\dot\phi + (\frac{\dot a}{ a} -
\frac{\ddot\phi_0}{\dot \phi_0}) \delta\phi$, while in the slow
roll approximation we have $\Gamma=\delta\dot\phi= a^{-1}
\pi^{y}$. We want to say that, upon quantization, the above
equation turns into
\begin{equation}\nabla^2 \Psi -
\mu  \Psi= s \langle\hat\Gamma\rangle. \label{main4}
\end{equation}
Before the collapse occurs, the expectation value on the right hand
side is zero. Let us now determine what happens after the collapse: To
this end, take the Fourier transform of (\ref{main4}) and rewrite it
as
\begin{equation}\label{modito}
\Psi_k(\eta)=\frac{s}{k^2+\mu}\langle\hat\Gamma_k\rangle_\Theta.
\label{Psi}
\end{equation}
Let us focus now on the slow roll approximation and compute the right
hand side, we note that $\delta\dot\phi=a^{-1}\py$ and hence
\begin{equation}
\delta\dot\phi_k=\frac{1}{a\sqrt{2}}[ g_k(\eta) (\ann_k^R
+i\ann_k^I)
 + \bar g_{k}(\eta) (\creR_{k}  + i\creI_{k})]
\end{equation}
For the expectation value we find
\begin{eqnarray}
\nonumber
\langle\Gamma_k\rangle_\Theta&=&\frac{\sqrt{\hbar L^3}}{a\sqrt{2}}
[ g_k(\eta) (  D^{R}_ke^{i\alpha^R_k}+i D^{R}_ke^{i\alpha^I_k})\\
 &+& \bar g_{k}(\eta) ( D^{R}_k e^{-i\alpha^R_k}+i
D^{R,I}_ke^{-i\alpha^I_k} )]\\
\nonumber
&=&{\sqrt{\hbar L^3k}}\frac{1}{2a} \times\\ && \left( D^R_k \cos(\alpha^R_k +k
\eta -\pi/2)
+ i D^I_k \cos (\alpha^I_k +k \eta -\pi/2)\right)\\
&=:&\sqrt{\hbar L^3 k}\frac{1}{2a}F(k). \label{F}
\end{eqnarray}
We note that we can write
\begin{equation}
\cos(\alpha_k +k \eta -\pi/2) =
\cos(\alpha_k+\gamma^c_k +  \Delta_k)
\end{equation}
where $ \Delta_k= k(\eta-\eta_k^c)= k\eta -z_k$. Then, using the
expressions (\ref{Col2}), and after a longer calculation, we find
\begin{equation}
F(k) = (1/2) [A_k (x^{R}_{k,1} +ix^{I}_{k,1}) + B_k (x^{R}_{k,2}
+ix^{I}_{k,2})]
\end{equation}
where
\begin{equation}  A_k =  \frac {\sqrt{ 1+z_k^2}} {z_k} \sin(\Delta_k) ; \qquad  B_k
=\cos (\Delta_k) + (1/z_k) \sin(\Delta_k).
\end{equation}
%-------------------------------------------------------------------
\section{Recovering the observational quantities.}
\label{sec_observation}
%-------------------------------------------------------------------
Now we must compare with the experimental results. We will only do
this to a certain approximation since we will disregard the changes to
dynamics that happen after reheating due to the transition to
standard (radiation dominated) evolution.

A crucial observation for what follows is to recognize the fact that
we can not measure $\Psi_k$
for each individual value of $k$. What we measure in fact is the
``Newtonian potential" on the surface of last scattering: $
\Psi(\eta_D,\vec{x}_D)$ which is a function of the coordinates on the
celestial two-sphere, i.e a function of two angles. From this we
extract
\begin{equation}
\a_{lm}=\int \Psi(\eta_D,\vec{x}_D) Y_{lm}^* d^2\Omega
\end{equation}
In fact the quantity that is measured is ${\Delta T \over T}
(\theta,\varphi)$ which is expressed as $\sum_{lm} \alpha_{lm}
Y_{l,m}(\theta,\varphi)$.  The angular variations of the temperature
is then identified with the corresponding variations in the
``Newtonian Potential" $ \Psi$, by the understanding that they are
the result of gravitational red-shift in the CMB photon frequency
$\nu$ so ${{\delta T}\over T}={{\delta \nu}\over {\nu}} = {{\delta (
    \sqrt{g_{00}})}\over {\sqrt{g_{00}}}} \approx\delta \Psi$.  Thus
we identify the theoretical expectation $\alpha_{lm}$ with the
observed quantity $\alpha^{obs}_{lm}$ The quantity that is
presented as the result of observations is $OB_l=l(l+1)C_l$ where
$C_l = (2l+1)^{-1}\sum_m |\alpha^{obs}_{lm}|^2 $. The observations
indicate that (ignoring the acoustic oscillations, which is anyway
an aspect that is not being considered in this work) the quantity
$OB_l$ is essentially independent of $l$, and this is interpreted
as a reflection of the ``scale invariance" of the primordial
spectrum of fluctuations.

To evaluate the quantity of interest we use (\ref{Psi}) and (\ref{F}) to
write
\begin{equation}
 \Psi(\eta,\vec{x})=\sum_{\vec k}\frac{s  U(k)} {k^2+\mu}\sqrt{\frac{\hbar
k}{L^3}}\frac{1}{2a}
 F(\vec{k})e^{i\vec{k}\cdot\vec{x}},
\label{Psi2}
\end{equation}
where we have added the factor $U(k)$ to represent the aspects of
the evolution of the quantity of interest associated with the
physics of period from reheating to decoupling, which includes
among others the acoustic oscillations of the plasma.  It is in
this expression that the justification for the use of statistics
becomes clear.  The quantity we are in fact considering is the
result of an ensemble of harmonic oscillators each one
contributing with a complex number to the sum, leading to what is
in effect a 2 dimensional random walk whose total displacement
corresponds to the observational quantity. To proceed further we
must evaluate the most likely value for such total displacement.
This we do with the help of the imaginary ensemble of universes,
and the identification of the most likely value with the ensemble
mean value. These two quantities are reasonably close in normal
circumstances, where the probability distribution has a single
local maximum (the global maximum), and is not pathological in
some other respect. Let us see how does this work in detail:

Using
$\vec{x}=R_D(\sin(\theta)\sin(\varphi),\sin(\theta)\cos(\varphi),
\cos(\theta))$ and standard results
connecting Fourier and spherical expansions
%\cite{Liddle and Lyth}
we obtain
\begin{eqnarray}
\alpha_{lm}&=&s\sqrt{\frac{\hbar}{L^3}}\frac{1}{2a} \sum_{\vec k}
\frac{ U(k)\sqrt{k}}{k^2+\mu} \int  F(\vec k)
e^{i\vec{k}\cdot\vec{x}}Y_{lm}
(\theta, \varphi) d^2\Omega\\
&=&s\sqrt{\frac{\hbar}{L^3}}\frac{1}{2a} \sum_{\vec
k}\frac{U(k)\sqrt{k}}{k^2+\mu} F(\vec k)  4 \pi i^l  j_l((|\vec k|
R_D) Y_{lm}(\hat k)\label{alm1}
\end{eqnarray}
where $\hat k$ indicates the direction of the vector $\vec k$. Now we
compute the expected magnitude of this quantity.  As a first step we
take the square of the quantity of interest:
\begin{equation}
|\alpha_{lm}|^2=s^2\frac{4\pi^2\hbar}{L^3}\frac{1}{a^2} \sum_{\vec
k,\vec{k'}}\frac{U(k)\sqrt{k}}{k^2+\mu}\frac{U(k')\sqrt{k'}}{k'^2+\mu}F(\vec
k)\cc{F(\vec{k'})}j_l(k R_D) j_l(k' R_D)Y_{lm}(\hat
k)Y_{lm}(\hat{k'}).
\end{equation}
Now we take its ensemble mean value.  The calculation of such mean
value will be simplified due to the fact that, as usual, the average
over the random variables will lead to an cancellation of the cross
terms. In our case we find
\begin{eqnarray}
\nonumber \langle F(\vec k)\cc{F(\vec{k'})}\rangle&=& A_k A_{k'} (
\langle x^{R}_{k,1} x^{R}_{k',1} \rangle +\langle x^{I}_{k,1}
x^{I}_{k',1}
\rangle)\\
&+&  B_k B_{k'} ( \langle x^{R}_{k,2} x^{R}_{k',2}
\rangle +\langle x^{I}_{k,2} x^{I}_{k',2} \rangle)
\end{eqnarray}
where we have made use of the independence among the four sets of
random variables $x^{ R}_{k,1}, x^{ I}_{k,1} ,x^{ R}_{k,2}, x^{
I}_{k,2}$. However we need to recall that within each set the
variables corresponding to $\vec k$ and $- \vec k$ are not
independent.  This will be reflected by setting $ \langle x^{
R}_{k}, x^{ R}_{k} \rangle = \delta_{k,k'} +\delta_{k,-k'}$ and $
\langle x^{ I}_{k}, x^{ I}_{k} \rangle = \delta_{k,k'}
-\delta_{k,-k'}$ in accordance with the discussion below equation
(\ref{NewCom1}).  Thus
\begin{equation}
\langle F(\vec k)\cc{F(\vec{k'})}\rangle =  ( A_k^2 + B_{k}^2)
\delta_{k,k'}
\end{equation}
Thus we arrive to the expression for the ensemble mean value which as
we said we will consider as a good approximation of the most likely
(M.L.) value:
\begin{equation}
|\alpha_{lm}|^2_{M. L.}= s^2\frac{4\pi^2\hbar}{L^3}\frac{1}{a^2}
\sum_{\vec k} ( A_k^2 + B_{k}^2) \frac{U(k)^2 k}{(k^2+\mu)^2}
j^2_l((|\vec k| R_D) |Y_{lm}(\hat k)|^2
\end{equation}
Now we write the sum as an integral by noting that the allowed values
of the components of $\vec k$ are separated by $\Delta k_i =2\pi/L$,
thus
\begin{equation}
|\alpha_{lm}|^2_{M. L.} =\frac{s^2  4\pi^2 \hbar}{a^2 L^{3}}
(L/2\pi)^3 \sum_{\vec k}\frac{( A_k^2 + B_{k}^2)  k
U(k)^2}{(k^2+\mu)^2}
 j^2_l((|\vec k| R_D) |Y_{lm}(\hat k)|^2 (\Delta k_i)^3
\end{equation}
\begin{equation}
= \frac{s^2   \hbar}{2 \pi a^2}\int \frac{ U(k)^2 C(k)
k}{(k^2+\mu)^2}    j^2_l((|\vec k| R_D) |Y_{lm}(\hat k)|^2 d^3k
\end{equation}
where $C(k)= A_k^2 + B_{k}^2$ is in fact a function of $k, z_k$,
and $\eta$.
\begin{equation}
=\frac{s^2  \hbar}{2 \pi a^2} \int \frac {U(k)^2
C(k)}{(k^2+\mu)^2} j^2_l((|\vec k| R_D)  k^3dk \label{alm4}
\end{equation}
where in the last equation we have made use of the normalization of
the spherical harmonics.  The last expression can be made more useful
by changing the variables of integration to $x =kR_D$ leading to
\begin{equation}
|\alpha_{lm}|^2_{M. L.}=\frac{s^2   \hbar}{2 \pi a^2} \int
\frac{U(x/R_D)^2 C(x/R_D)}{(x^2+\mu R_D^2)^2}    j^2_l(x) x^3 dx
\label{alm5}
\end{equation}
In the exponential expansion regime where $\mu$ vanishes, and in
the limit $z_k\to -\infty$ where $C=1$, and taking for simplicity
$U(k) =U_0$ to be independent of $k$, (neglecting for instance the
physics that gives rise to the acoustic peaks), we find:
\begin{equation}
|\alpha_{lm}|^2_{M. L.}=\frac{s^2  U_0^2  \hbar} {2 \pi a^2}
I_1(l)
\end{equation}
where $ I_n (l) =\int x^{-n } j^2_l(x) dx$. For $n=1$ we have $I_1 (l)
={\frac{\pi}{l(l+1)}}$ so
 \begin{equation}
 |\alpha_{lm}|^2_{M. L.}=\frac{s^2  U_0^2  \hbar} {2  a^2}
\frac{1}{l(l+1)} .
\end{equation} Now,  since this does not depend on $m$ it
is clear that the expectation of $C_l = (2l+1)^{-1}\sum_m
|\alpha_{lm}|^2 $ is just $|\alpha_{lm}|^2$ and thus the
observational quantity $OB_l=l(l+1)C_l =\frac{s^2 U_0^2 \hbar}{2
a^2} $ independent of $l$ and in agreement with the scale
invariant spectrum obtained in ordinary treatments and in the
observational studies.  Now let us look at the predicted value for
the observational quantity $OB_l$. Using the expression $s= 4\pi
G\dot\phi_0$, the equation of motion for the scalar field in the
background in the slow roll approximation ($\dot\phi =-\frac{
a^3}{3\dot a} V'$ where $ V'=\frac{\partial
  V}{\partial \phi}$), and the first of Einstein's equations, in the
background which gives $3(\dot a)^2 = 8\pi G a^4 V(\phi_0)$, we find,
\begin{equation}
OB_l= (\pi/6) G\hbar \frac{(V')^2}{V} U_0^2 =
(\pi/3)\epsilon (V/M_{Pl}^4) U_0^2
\end{equation}
where in the last equality we have used the standard definition of
the slow roll parameter $\epsilon= (1/2) M_{Pl}^2 (V'/V)^2$, and
$G\hbar =M_{Pl}^{-2}$.  Note that if one could avoid $U$ from
becoming too large during reheating, which is another aspect of
the problem where the complications of the relevant physics might
leave room for a departure from the standard results, the quantity
of interest would be proportional to the small number $\epsilon$,
a possibility that is not discussed in the standard treatments, so
in this case we could get rid of the "fine tuning problem" for the
inflationary potential (i.e. even if $ V\sim M_{Pl}^4$, the
temperature fluctuations in the CMB would be expected to be
small).

{ Furthermore we note, as can be seen from  inspection of
(\ref{alm5}), that if the evolution of the scale factor deviates
slightly from the one given in (\ref{expansion}) in such a way that
$\mu>0$, the effect would be to decrease the value of $OB_l$ relative
to the standard prediction of the scale invariant spectrum, for the
small values of $l$. Such effect\footnote{It is however important to
recall that, as we have indicated, all that can be extracted from the
theory is an estimate of the ``most likely" value" of $\alpha_{lm}$,
as it results from a sort of random walk of the individual collapses.
What is more, the likelihood of a deviation  has to do with the
number of steps -- i.e. individual collapsing  modes -- that
effectively contribute to the random walk. This number in turn would
be depend, as can be clearly seen in the discussion above
(\ref{APPalpha-lm}) of Appendix \ref{app_spherical} -- on
 global features of the universe.  It is clear that the likelihood  of a deviation
grows with the lowering  of the values of $l$.} has been reported
\cite{Low-l-anomaly}, as we mentioned in the introduction, and has
been the subject of quite some interest lately.  More details
about these data can be found in \cite{Boomerang-Maxima-WMAP}.
Moreover we note that by a detailed analysis of this sort, one
could in principle extract information about the deviations form
the standard exponential expansion normally associated with
inflation from the observational data.

Now let us focus on the} effect of the finite value of time of
collapse $\eta^c_k$, that is we consider the general functional
form of $C(k)$:
\begin{equation}
C(k)=1+ (2/ z_k^2) \sin (\Delta_k)^2 + (1/z_k)\sin (2\Delta_k)
\label{ExpCk}
\end{equation}
where we recall that $\Delta_k= k \eta -z_k$ with $ z_k =\eta_k^c k$.
The first thing we note is that in order to get a reasonable spectrum
there seems to be only one simple option: That $z_k$ is essentially
independent of $k$ -- that is the time of collapse of the different
modes should depend on the mode's frequency according to
$\eta_k^c=z/k$. { Recall that the standard answer would correspond to
$C(k)=1$. This can be obtained here
  in the limit of very early collapse$ z_k \to - \infty$, or if one takes the
time of collapse equal to the moment of observation\footnote{One possibility that seems
to emerge from the analysis this far is for such single event to
occur at the time of
 decoupling so $\Delta_k=0$.  For a
possible interpretation of this see next section.}. Such  possibility will not be considered further at this point.    In  the case of a finite value for the time of collapse
 what we seek is for $C(k)$ to be independent of $k$.
 In such case,  the collapse  would need  be associated with a mechanism that affects each
mode at the appropriate time.
 This is a
remarkable conclusion which provides relevant information about
whatever the mechanism of collapse is. }
This is
however not enough to ensure that the resulting spectrum coincides
 exactly with the one obtained in the observations. For that we need to
ensure that the other relevant parameter $\eta k$ does not
introduce a significant $k$ dependence. However in this case
$\eta$ which refers to the conformal observation time (the time at
which the observed photons are emitted, corresponding in the
realistic case to the time of decoupling), is of course the same
for all modes. Then unless the effect of the quantity $\eta k$
becomes negligible we will not be able to account for an exactly
scale invariant spectrum.
%---------------------------------------------------------------
\section{Further ideas}
\label{sec_further}
%---------------------------------------------------------------
In this section we have collected some further ideas relating to
the transition to classicality and inhomogeneity: In the following
subsection we present a point of view that is to a large extent
inspired in the spirit of the Many Universes point of view (see
Sec. \ref{sec_other}), together with the idea that missing
ingredient would be provided by Quantum Gravity. In Subsection
\ref{sec_penrose} we discuss how a mechanism that triggers
collapse based on the gravitational interaction energy of quantum
mechanical alternatives -- as envisioned by Penrose -- can be
incorporated within the present formalism.
%----------------------------------------------------------------
\subsection{A special role of gravity in the Many Universes
  interpretation?}
%---------------------------------------------------------------
A possible view contemplates a quantum evolution all the way up to
the time of observation, today. In this view the gravitational
field is all the time related to the matter through the field
equation (\ref{eq_fieldeq}) which is imposed as an operator equation
(no need to any semiclassical approximation in the linearized regime).

Consider, for an analogy, a Stern-Gerlach experiment where we have an
incoming electron propagating along the $x$-direction in a state
$|y,+\!\!>$ (spin up in the $y$-direction) with the magnetic field of
the apparatus oriented in the $z$-direction (corresponding to a
specific quantum state of the apparatus). We also introduce an
observer which is treated quantum mechanically. In this case, the
state of the apparatus selects a preferred basis where we can
describe the quantum evolution. The Schr{\"{o}}dinger evolution of
the electron-apparatus-observer quantum system is such that, after
the electron goes though the apparatus, the state of the system
evolves into
\[|\psi>=(|z,+\!\!>\otimes |x^+\!\!>\otimes|obs.+\!\!>)+
(|z,-\!\!>\otimes |x^-\!\!>\otimes|obs.-\!\!>),\] where
$|x^{\pm}\!\!>$ represent the two possible trajectories (up and
down) of the electron passing through the Stern-Gerlach apparatus,
and $|obs.\pm\!\!>$ the associated two states of the observer's
mind.  The choice of this basis allows the interpretation of this
final state where two alternative ``worlds'' are described by the
state vector. In each one of these worlds the observer sees the
electron going up or the electron going down respectively. This
interpretation seems unambiguous due to the special role of the
apparatus in selecting a preferred basis in the Hilbert space of
the joint system.

If we what to interpret the result of the quantum evolution of the
inflaton field during inflation we are confronted with what seems
to be a similar situation. On the one hand, the observations of
the CMB are directly related to the fluctuations of the geometry
at the time of decoupling. Let us for the moment assume that the
analog of $x^{\pm}$ is the value of the Newtonian potential
$\psi$. To linearized order quantum Einstein's equations are given
by the operator version of (\ref{eq_fieldeq}) which becomes:
\[\nabla^2 \hat\psi=\frac{4\pi G\dot \phi_0}{a}\hat \pi^{(y)},\]
It is clear that this equation connects the eigenbasis of the
Newtonian potential with the eigenbasis of the momentum  of the
scalar field. At each instant of time $\eta$ the wave function of
the Newtonian potential $\Psi(\psi_k)$ of each mode is related by
the previous constraint to the (squeezed) state resulting from the
evolution of the vacuum wave function of $\Psi(\pi^{(y)})$
associated to the scalar field. The state of the universe is thus
given by $\Psi(\psi_k)\otimes\Psi(\pi^{(y)})$ which can be written
as a linear superposition of eigenstates of the Newtonian
potential and the scalar field momentum. Due to the linearity of
quantum mechanics, the evolution of the full state (which is
always homogeneous and isotropic) can be analyzed by evolving each
component separately. Every one of these alternatives will
unitarily evolve in such a way that inhomogeneities grow, form
structure, produce galaxies, and finally observers, yet the full
state remains homogeneous and isotropic. In analogy to the
Stern-Gerlach experiment we can express the final quantum state of
the universe as a superposition of states with its own observers.
One then  takes the view, that what quantum mechanics predicts
about the universe is the probabilistic distribution of these
alternatives. We can use the above formalism to make predictions
about the inhomogeneities in our own observed world by using
definition (\ref{alm1}) in order to estimate the most likely value
of $|\alpha_{lm}|$ as $\sqrt{<\hat \alpha_{lm}^2>}$ which also
leads to (\ref{eq_semiq}) with $C(k)=1$.

Of course there are certain issues with this viewpoint:

{\em \noindent What selects the basis?}
Unlike the Stern-Gerlach situation there does not seem to be an
obvious physical input that would lead to a preferred basis where
to make the decomposition. One could take the view that
 the basis is the one of the eigenstates of the gravitational
 field. After all when we observe the CMB
 we do measure $\psi$; therefore, as in standard situations in QM
 we select the basis when we decide what to measure. For example
 if one measures the spin of a particle in the z-direction one should
 write the state in its eigenbasis to make any prediction.
This fails to address the question of how we are there to make the
measurement in the first place. Only in a very specific basis
would matter evolve to form observers. It seems that here we are
again confronted to some unknown mechanism selecting this basis.
Gravity plays a central role in the mechanisms that allows
structure to grow and produce the conditions that lead to the
existence of observers. This suggests that the  gravitational
realm is perhaps the place where this new element should be found.

{\em \noindent What are the limits to which this position can be
taken?} We should start by noting that in the position exposed
above the full quantum state represents a set of coexistent
possibilities. First of all the separation of alternatives cannot
be made in a strict eigenbasis of any field as the conjugate
quantity would produce large departures from what we would call
the near classical evolution. The exact nature of the states that
constitute such set of non interfering alternatives should  be
provided by the new element we are calling upon.  Furthermore this
element should not prevent normal quantum mechanical interference
in standard situations. Some sort of threshold is needed.
Presumably at the very early stages of the evolution such
threshold would not have been reached, and thus the situation
would have been described as a fully interfering quantum state,
the one corresponding to the scalar field vacuum. This seems to
take us back to a similar sort of scenario as the one investigated
in this paper.

%---------------------------------------------------------
\subsection{A `Penrose mechanism' for collapse}
\label{sec_penrose}
%--------------------------------------------------------
Penrose has for a long time advocated that the collapse of quantum
mechanical wave functions might be a dynamical process independent
of observation, and that the underlying mechanism might be related
to gravitational interaction. More precisely, according to this
suggestion, collapse into one of two quantum mechanical
alternatives would take place when the gravitational interaction
energy between the alternatives exceeds a certain threshold. In
fact, much of the initial motivation for the present work came
from Penrose's ideas and his questions regarding the quantum
history of the universe.

A very naive realization of Penrose's ideas in the present setting
could be obtained as follows: Each mode would collapse by the
action of the gravitational interaction between its own possible
realization. In our case one could estimate the interaction energy
$E_I(k,\eta)$ by considering two representatives of the possible
collapsed states on opposite sides of the Gaussian associated with
the vacuum. Let us interpret $\Psi$ literally as the Newtonian
potential and consequently the right hand side of equation
(\ref{main2}) as the associated matter density. Then we would have
\be E_I(\eta)=\int \Psi^{(1)}(x,\eta) \rho^{(2)}(x,\eta)dV = a^3
\int \Psi^{(1)}(x,\eta) \rho^{(2)}(x,\eta) d^3x \ee which when
applied to a single mode becomes: \be E(\eta)= (a^3/L^6)  \Psi_{
k}^{(1)}( \eta) \rho^{(2)}_{k} (\eta) \int d^3x = (a^3/L^3)
\Psi^{(1)}_{ k}( \eta) \rho^{(2)}_{k} (\eta) \ee where $(1),(2)$
refer to the two different realizations chosen, and $\rho_{k}
=\dot\phi_0 \Gamma_k$, $\Psi_{ k} = ( s/k^2) \Gamma_k$,
 where $\Gamma_k =\pi^y_k/a$ and $s= 4\pi G\dot\phi_0$.
From equation (\ref{momentito}) we get
 $|<\Gamma_k > |^2 = \hbar k L^3 (1/2a)^2$.
Then \be E_I(k,\eta) = ( \pi/4) (a/k) \hbar G (\dot\phi_0)^2. \ee
In accordance to Penrose's ideas the collapse would take place
when this energy reaches the `one-graviton' level, namely when
\[E_I(k,\eta)=M_p,\] where $M_p$ is the
Planck mass. Using equations (\ref{modito}) and (\ref{momentito})
one gets \be z_k=\frac{\pi \hbar G \dot \phi_0^2}{H_I M_p}. \ee So
$z_k$ is independent of $k$ which according to equation
(\ref{ExpCk}) leads to a roughly  scale invariant spectrum of fluctuations in
accordance with observations.

A more detailed exposition of this scenario and its ramifications is
however beyond the scope of the present paper and will be taken up
elsewhere.

%--------------------------------------------------------------------
\section{ Discussion }
\label{sec_disc}
%------------------------------------------------------------------
We have discussed the problematic part of the standard analysis
that is supposed to predict the primordial spectrum of
fluctuations responsible for the deviation of our universe from
perfect homogeneity and isotropy and in particular for the
eventual evolution of galaxies, stars, and our own. We have argued
that there is an essential element that is missing in existing
proposals. We have argued that the missing element must contain
some new physics. We have considered this issue following the line
of thought exposed by Penrose, that such new physics might be tied
to some quantum aspect of gravitation, and we have employed this
idea in what we called the collapse hypothesis, which is reflected
concretely in our model in the fact that we take the Newtonian
potential to couple to expectation values of the quantum matter
degrees of freedom, and have allowed such expectation values to
``jump'' in association with the collapse process in a particular
set of states. It should thus be emphasized that this can be
justified only if we declare that gravitation is, at the quantum
level, profoundly different from other degrees of freedom as only
such posture would justify the different treatment awarded to the
gravitational and the scalar sectors in the present work. We have
shown that a relatively simple proposal concerning a collapse of
the wave function induced by some unknown mechanism, possibly tied
to Quantum Gravity, can account in a transparent way for the scale
invariant spectrum that seems to fit very well with the
observations.

In Appendix \ref{app_spherical} we have shown a degree of
robustness of our approach by redoing the analysis starting from
the onset with a spherical mode expansion of fields, and assuming
that the collapse directly affects these spherical modes. This
fact is not entirely trivial, because as we have argued before,
the outcome of a collapse is associated with the quantity that is
selected by the ``measurement". Thus the exact nature of the
ensemble of collapsed harmonic oscillators that constitute our
field depends on the modes one chooses to define the collapsed
state. However the specific statistical properties of the ensemble
that one is examining turn out to be the same in the two
situations that we have considered.
%
%i) that where the collapsed state is associated with the Cartesian
%modes and ii) that in which it is associated with the spherical
%modes.
%
It is not unnatural to expect this result to be generic if one
thinks of it in analogy with an EPR type of situation: The
statistical properties determined by an experimentalist on one end
of the set up are independent of the axis
  of collapse selected by the choice of polarization used by
  the experimentalist in the other end, whose measurements could be
  thought to trigger the collapse in that case. However in our
  situation the possibility of some differences in other more subtle
  statistical aspects of the ensemble can not be so easily ruled out
  as we do have access to the system on which the measurements must
  have occurred (i.e. in contrast with the standard considerations for
  an experimentalist on one end of the EPR setup, we are not denied, in
  principle, access to regions of the universe that might contain
  relevant information about correlations).  Note on the other hand
  that access to such correlations might be ruled out
  using arguments similar to those employed by Peres to
  conclude that one can not experimentally determine the quantum state of
  a single photon, if one does not have access to the information of
  how it was prepared \cite{Asher}.  One place differences can
  conceivably arise is in the correlations among the $\alpha_{lm}$
  for different $m$ and fixed $l$. These issues are of course in need of
  further investigation.

Furthermore we have shown that, in this scheme, the resulting
amplitude for the fluctuations could become naturally small
provided the physics at and after reheating does not introduce
  unwanted amplifications, a possibility that could not be easily uncovered in
  the standard treatments, probably due to the fact that they do not
  allow for a transparent picture of the time at which the departure
  from homogeneity and isotropy starts, and is subjected to different
  physical regimes of evolution. Thus our proposal opens the door for a study dedicated to
  eliminating the extreme fine tuning that is necessary in previous
  treatments.
Given that our motivation had nothing to do with this quantitative
issues we view this result as another indicative of the promising
value of our approach. Furthermore we have indicated how a small
departure from the exponential expansion usually associated with
inflation could explain the observed decrease in the amplitude of
fluctuations on large angular scales.

{ It is  clear {\it the scheme is in principle susceptible to
experimental
  exploration} as is  shown by the non trivial form of the
  function $C(k)$ ( equation (\ref{ExpCk})) which contains information about
  the collapse times and modes, and which is an input to equation (\ref{alm5}),
 which gives the quantities that  are to be compared with experiment.
 In particular in appendix \ref{app_alternative} we show
  explicitly that a small change in the collapse scheme leads to
  differences in the form of this function. Recall that the ``standard answer" would correspond to
  $C\equiv 1$, a result that is not trivial to obtain in this scheme as discussed in the end of section
  \ref{sec_observation}.}

In addition our treatment reveals the various different
statistical aspects that are at hand: First there is the
statistical averaging over the modes $k$ that contributes to a
specific $a_{lm}$. Then one has the statistics that is associated
with their average over $m$ which leads to the observational
quantity $C_l$ or $OB_{l}$ in terms of which the experimental
results are often exhibited. This clear disentanglement opens the
door for more elaborated statistical analysis of the data. In
particular we note that the disappearance of all information
concerning the size of the sphere of last scattering and the
effective region of the universe that contains the modes that
contribute to the observations is particular to the examined
values of the angular momentum and to the slow role approximation.
That is, the cancellations of the quantities $R_D$, and $L$, from
our final results would not occur if either we do not neglect the
terms proportional to $\mu$ in (\ref{alm5}), or if we consider
small deviations from slow roll approximation. This would lead to
the appearance of $\delta \phi$ in (\ref{Psi}), and thus the
source term $\Gamma$ for the Newtonian potential $\Psi$ would
contain not only the term proportional to $\sqrt{k}$ that comes
from $g_k(\eta)$ but also terms proportional to $k^{-1/2}$ and
$k^{-3/2}$ coming from $y_k(\eta)$.  This would result in the
appearance of different powers of $k$ in equations (\ref{F}),
(\ref{Psi2}), (\ref{alm1}), and (\ref{alm4}) which would lead to a
scale dependent contribution to the spectrum. If the observations
could be improved substantially it might be possible to see these
effects. One potential source of information in this respect is in
some sense already available: One could look at the observed
values of the different $\alpha_{lm}$ for each fixed value of $l$
instead of concentrating in their average $C_{l}$ and thus extract
from the scatter in these quantities information about the
effective number of steps in the two dimensional random walks,
that is the effective number of $\vec k$'s that contribute to
their value. This in turn would provide information about the
``effective size of the Universe".

Finally we acknowledge we have said almost nothing about the
physical process that triggers the collapse of the wave function
(except the example considered  briefly in section
\ref{sec_penrose} that we take at this point to be a simple
illustrative model). We have done this consciously because our aim
in this work was to point out the difficulties with the standard
views on the issue, and to illustrate the kind of effect we need
the new physics to bring about. It is clear for instance,  that
during the collapse process the semi-classical Einstein equations
(\ref{eq_semi}) can not be satisfied. We have in mind however that
this is some approximation to a more complete description
including terms tied to quantum gravity and to the
 mechanism that triggers the collapse. Instead of
 equation (\ref{eq_semi}), we would be considering
\begin{equation}
\label{eq_semiq}
G_{ab}+Q_{ab}=8 \pi G \expec{T_{ab}},
\end{equation}
where $Q_{ab}$ represents the back reaction of the geometry to the
changes of the expectation values of the energy momentum tensor
associated with the collapse of the quantum state of the field.
 This mechanism must
therefore be such that the manipulation made within the
semi-classical treatment of Einstein's gravity would be justified,
except when the collapse takes place, the points at which the new
terms would become important. In other words we need
 Einstein's  equation to be modified by a
term, that reflects the mechanism of collapse and that is
responsible   for important changes only during the collapse
itself. It is our hope that  the examination of this and the other
requirements we have found so far for the collapse mechanism would
be useful guides in constructing specific
 models of the new physics behind it.  We should
stress again that our ideas in this regard are strongly
influenced by Penrose's proposals that some unknown aspect of
Quantum Gravity might be at play. We have briefly  considered one simple
example of these ideas in \ref{sec_penrose},
 and clearly  much more detailed  analysis of these issues is required.
In this regard, and following a different line of thought,  it is worth
pointing out that the quantum
uncertainties in the sources of the Newtonian potential could be
thought as inducing the kind of quantum superposition of different
geometries that Penrose associates with the mechanism that
triggers the collapse.
 In this context we note that the quantum
uncertainty in $\delta\dot\phi$ --which is the one that leads to
observed spectrum -- behaves as $1/a$ while  the quantum
uncertainty in $\delta\phi$  decreases at a much slower rate.  The
volume over which these uncertainties would in principle affect
the Newtonian Potential grows as $a^3$ so their effect would in
principle grow with the universe's size, and thus in the spirit of
Penrose's ideas, the universe would, as it evolves, be approaching
from below the threshold where the superposition of geometries
would lead to a spontaneous collapse of the wave function. On the
other hand we point out that $\delta\phi$ disappeared in our
treatment as a source of the Newtonian potential  due to the slow
roll approximation. Thus, one might be tempted to think that
higher order  perturbative gravitational effects associated with
the latter could be the trigger, in the spirit of Penrose's ideas,
of the quantum gravitationally induced collapse. These issues will
be the subject of further investigations.

We end by noting a paradoxical aspect of the situation in our
field of study:  On the one hand there is an almost frenetic
search for any form of experimental manifestations of any
conceivable aspect of quantum gravity, while on the other hand,
when faced with as clear an arena for these studies, as the one we
have treated in this work, the prevailing attitude seems to be to
hide the mysteries under the rug and declare that everything is
fine. It is our hope that this paper contributes to changing this
situation.

\section*{Acknowledgments}

\noindent We gratefully acknowledge very useful discussions with
Roger Penrose, Abhay Ashtekar,  James Bjorken, Carlo Rovelli,
Michael Reisenberger.

This work was  supported in part by DGAPA-UNAM IN108103 and CONACyT
43914-F grants, by NSF grant PHY-00-90091, and by the
Eberly research funds of Penn State. HS gratefully acknowledges
funding by the Max-Planck Society through the MPI for Gravitational Physics.
\section*{Appendix}
\begin{appendix}
%-------------------------------------------------------------------
\section{Spherical expansion}
\label{app_spherical}
%-------------------------------------------------------------------
The symmetry of the situation and the physical application makes it
more natural to use spherical coordinates on the spatial slices, and
spherical harmonics as a basis for the one particle space. We start
again with the equations for the field $y=a\delta \phi$,
\begin{equation}
\label{equ1}
 \ddot y-\left(\nabla^2 +\frac{\ddot a}{a}\right)y=0,
\end{equation}
where we now write the Laplacian on Euclidean three space as
\begin{equation}
\nabla^2 \phi= r^{-2}
\partial_r (r^2 \partial_r\phi) -r^{-2} J^2 \phi,
\end{equation}
where $J$ stands for the angular momentum differential operator. In
order to quantize the auxiliary field $y$, we consider now the field
in a spherical box of radius $R$ and require the field to vanish at
its boundary.
%
%
%The symmetry of the situation and the physical application makes it
%more natural to use the following coordinatization;
%\begin{equation}
%ds^2=a^2(\eta)[d\eta^2-(dr^2+r^2
% d\Omega^2)],
%\end{equation}
%In these coordinates, the field equation is again
%\begin{equation}
%\delta \ddot \phi-\nabla^2 \delta \phi +2\frac{\dot a}{a}\delta
%\dot\phi=0
%\end{equation}
%where dots denote derivatives with respect to $\eta$ and $\nabla^2$ is
%the Laplacian on Euclidean three space in the spherical coordinates
%(whose metric is $dr^2+r^2 d\Omega^2$) and can be written as
%\begin{equation}
%\nabla^2 \phi= r^{-2}
%\partial_r (r^2 \partial_r\phi) -r^{-2} J^2 \phi
%\end{equation}
%where $J$ stands for the angular momentum differential operator.
%Again equation can be further simplified by going over from $\delta
%\phi$ to an auxiliary field $y=a\delta \phi$.  In the resulting
%equation for $y$, and again no term with a first derivative of the
%field anymore,
%\begin{equation}
%\label{equ1}
% \ddot y-\left(\nabla^2 +\frac{\ddot a}{a}\right)y=0,
%\end{equation}
%In order to quantize the auxiliary field $y$, we consider now the
%field in a spherical box of radius $R$ and require the field to vanish
%at its boundary.  We can think for instance in the case of a closed
%universe so that the boundary is in fact a point and thus the non
%trivial angular dependence in which we will be interested requires for
%the single valuedness of the field that it vanishes at such point.
%
The standard separation of variables leads us to write the field in
terms of the modes
\begin{equation}
U_{klm}=y_k(\eta)f_{kl}(r)Y_{lm}(\theta\varphi)
\end{equation}
where the $Y_{lm}$ are the standard spherical harmonics and where
the combination $f_{kl}(r)Y_{lm}(\theta\varphi)$ are eigenmodes of
the Laplacian with eigenvalue $-k^2$. In particular we have
$f_{kl}=c_{kl} j_l(kr)$, where $j_l$ are the spherical Bessel
functions, and the normalization is such that
\begin{equation}
\int_0^R r^2dr
\int d\Omega |f_{kl}(r)Y_{lm}(\theta\varphi)|^2=1.
\end{equation}
That is $c_{kl}= k^{3/2} [\int_0^{x_i^{(l)}} (x j_l(x))^2)
dx]^{-1/2}$ where $x_i^{(l)}$ corresponds to the $i^{th}$ zero of
$j_l(x)$, and where the allowed values of $k$ for a given $l$ are
those that satisfy $kR=x_i^{(l)}$. The $y_k(\eta)$ have to satisfy
the same equation as before, and thus we can quantize the field
$y$ as
\begin{equation}
   \y(\eta,r, \theta, \varphi)= \sum_{ k, l,m}
\left[\ann_{k,l,m}y_k(\eta)f_{kl}(r)Y_{lm}(\theta\varphi) + \cre_{k,
l,m}\bar y_k (\eta)f_{kl}(r)\bar Y_{lm}(\theta\varphi) \right],
\end{equation}
where the sum is over the allowed values of $k$ for each $l$ and
as usual $m \in \lbrace -l,...,l\rbrace$, and we have again chosen
the positive frequency solutions given by (\ref{Sol-y}) for the
$y_k$. The canonical conjugate to $y$ can be written in a similar
fashion, with $g_k$ taking the place of $y_k$.
%
%
%Now we have,
%\begin{equation}
%   \ddot y_k+\left(k^2 -\frac{\ddot a}{a}\right)y_k=0,
%\end{equation}
%which is the same equation we found before, thus again we take
%$y_k(\eta)$ the positive frequency solution given by equation
%(\ref{Sol-y}). Now we decompose the real classical field $y$ into
%these modes and then promote it to a quantum field writing:
%\begin{equation}
%   \y(\eta,r, \theta, \varphi)= \sum_{ k, l,m}
%\left[\ann_{k,l,m}y_k(\eta)f_{kl}(r)Y_{lm}(\theta\varphi) +
%\cre_{k, l,m}\bar y_k (\eta)f_{kl}(r)\bar Y_{lm}(\theta\varphi)
%\right],
%\end{equation}
%where the sum is over the allowed values of $k$ for each $l$ and as
%usual $m \in \lbrace -l,...,l\rbrace$. Similarly the canonical
%conjugate to $y$ can be written as;
%\begin{equation}
%\py(\eta,r, \theta, \varphi)= \sum_{ k, l,m}
%\left[\ann_{k,l,m}g_k (\eta)f_{kl}(r)Y_{lm}(\theta\varphi) +
%\cre_{k, l,m}\bar g_k (\eta)f_{kl}(r)\bar Y_{lm}(\theta\varphi)
%\right],
%\end{equation}
%where again
%\begin{equation}
%  g_k=\dot y_k-\frac{\dot a}{a}y_k.
%\end{equation}
The standard commutation relations are equivalent to
\begin{equation}
[\ann_{k,l,m},\cre_{k', l',m'}]=\hbar
\delta_{k,k'}\delta_{l,l'}\delta_{m,m'}.
\end{equation}
Our definition of the components of a function $F$ in terms of the
spherical harmonics will be $F_{k,l,m}:=\int d^3x (f_{kl}Y_{lm})^*
F$. For the field modes we find
\begin{equation}
\y_{k,l,m}(\eta) =y_k(\eta)\ann_{k,l,m} +\cc{y_k}(\eta)\cre_{k,l,-m},
\end{equation}
and express their real and imaginary parts as
\begin{equation}
\label{eq_decomp1} \y_{k,l,m}(\eta)^R:= (1/2)[ \y_{k,l,m}(\eta)
+\y_{k,l,m}(\eta)^{\dagger}] =\frac{1}{\sqrt{2}} [
y_k(\eta)\ann_{k,l,m}^R +\cc{y_k}(\eta)\creR_{k,l,m}]
\end{equation}
\begin{equation}
\label{eq_decomp2}
 \y_{k,l,m}(\eta)^I:= (1/2i)[ \y_{k,l,m}(\eta)
+\y_{k,l,m}(\eta)^{\dagger}] =\frac{-i}{\sqrt{2}} [
y_k(\eta)\ann_{k,l,m}^I -\cc{y_k}(\eta)\creI_{k,l,m}].
\end{equation}
The real and imaginary components of the annihilation operators are
defined by
\begin{equation}
   \ann_{k,l,m}^R :=\frac{1}{\sqrt{2}} [\ann_{k,l,m} +\ann_{k,l,m}] \
\ann_{k,l,m}^I := \frac{-i}{\sqrt{2}} [\ann_{k,l,m} -\ann_{k,l,m}] \
\end{equation}
and conjugation. Consequently
\begin{equation}
[\ann_{k,l,m}^R, \creR_{k',l',m'}] = \hbar \delta_{k,k'}\delta_{l,l'}
(\delta_{m,m'} + \delta_{m,m'})
\end{equation}
\begin{equation}
[\ann_{k,l,m}^I, \creI_{k',l',m'}] = \hbar \delta_{k,k'}\delta_{l,l'}
(\delta_{m,m'} - \delta_{m,m'})
\end{equation}
\begin{equation}
[\ann_{k,l,m}^R, \creI_{k',l',m'}] =  [\ann_{k,l,m}^I,
\creR_{k',l',m'}]=0.
\end{equation}
The modes of $\py$ can be decomposed in the same way. The only
thing that changes in comparison to
(\ref{eq_decomp1},\ref{eq_decomp2}) is that $y_k$ gets replaced by
$g_k$.
Now we consider the quantity of interest that is determined by the
following quantity,
\begin{equation}
d^{R,I}_{k,l,m}= < \ann_{k,l,m}^{R,I}>_{\Theta}
:=\sqrt{\hbar}D^{R,I}_{k,l,m} e^{i\alpha^{R,I}_{k,l,m}}
\end{equation}
These quantities which as before associated with the collapsed state
expectation value of the field and momentum, which in our model for
the collapse are determined by the relation at the time of collapse
\begin{eqnarray}
\expec{\y_{k,l,m}^{R,I}(\eta^c_k)}_\Theta&=&x^{R,I}_{k,l,m,1}\sqrt{\fluc
{\yRI_{k,l,m}}_0}=x^{R,I}_{k,l,m,1}|y_{k}(\eta^c_k)|\sqrt{\hbar/2 },\\
\expec{\pyRI_{k,l,m}(\eta^c_k)}_\Theta&=&x^{R,I}_{k,l,m,2}
\sqrt{\fluc{\pyRI_{k,l,m}}_0}=x^{R,I}_{k,l,m,2}|g_k(\eta^c_k)|\sqrt{\hbar/2},
\end{eqnarray}
where as before we have for any time at or after the collapse,
\begin{eqnarray}
\expec{\y_{k,l,m}^{R,I}(\eta_k)}_\Theta&=&\sqrt{2} \Re
(y_{k}(\eta_k)
d^{R,I}_{k,l,m}),\\
\expec{\pyRI_{k,l,m}(\eta^c_k)}_\Theta&=&\sqrt{2} \Re
(g_{k}(\eta_k) d^{R,I}_{k,l,m})
\end{eqnarray}
Again in terms of the phases $\beta_k=\arg(y_k )$,
$\gamma_k=\arg(g_k)$, where the last two refer to quantities
evaluated at the collapse time $\eta^c_k$, the above equations can be
written
\begin{equation}
D^{R,I}_{k,l,m}\cos(\alpha^{R,I}_k+\beta^c_k)=\frac{1}{2}x^{R,I}_{k,l,m,
1}, \qquad
D^{R,I}_{k,l,m}\cos(\alpha^{R,I}_k+\gamma^c_k)=\frac{1}{2}x^{R,I}_{k,l,m
,2}.
\label{Col2a}
\end{equation}
To connect to observations, we again use the relation $\nabla^2 \Psi
- \mu  \Psi= s  \Gamma$. Decomposing both sides into spherical
harmonics, we find
\begin{equation}
\Psi_{klm}(\eta)= -\frac{s U(k)}{(k^2+\mu)}\Gamma_{klm}(\eta)
\end{equation}
where, once more, the factor $U(k)$ accounts for the physical
evolution from reheating to decoupling. The right hand side is
connected to the scalar field via, $\Gamma= \frac{1}{a} \pi,$ so in
our context
\begin{equation}
\Gamma_{k,l,m}(\eta)= \frac{1}{a} \expec{
\py_{k,l,m}(\eta)}_{\Theta}.
\end{equation}
The observational quantity $\alpha_{lm}$ can be expressed as
\begin{equation}
\alpha_{lm}=\int \Psi(\eta_D,\vec{x}_D) Y_{lm}^* d^2\Omega =\sum_{k}
\Psi_{klm}(\eta_D)
 f_{kl}(R_D).
\end{equation}
Again we see that this quantity is a sum of the complex quantities
associated with the collapse and thus can be viewed as a two
dimensional random walk with variable step size.  We resort once
more to the mathematical trick of identifying the most likely
value of the this magnitude with the mean value of an ensemble of
identical instances. This corresponds to identifying the most
likely value of the total displacement of the random walk with its
ensemble average. Thus we write
\begin{equation}
|\alpha_{lm}|^2_{Most Likely}=<<|\alpha_{lm}|^2>> =<<|\sum_{k}
\Psi_{klm}(\eta_D)|^2>>,
\end{equation}
\begin{equation}
=\sum_{k,k'}\frac{s^2
U(k)U(k')}{(k^2+\mu)(k'^2+\mu)}<<\Gamma_{klm}(\eta_D)\Gamma_{k'lm}(
\eta_D)>> f_{kl}(R_D) f_{k'l}(R_D)
\end{equation}
Now we evaluate this ensemble average noting that
\begin{eqnarray}
\Gamma_{klm}(\eta) &=&(1/a) \lbrace
\expec{\pyI_{k,l,m}(\eta)}_\Theta
+i\expec{\pyI_{k,l,m}(\eta)}_\Theta\rbrace \\
&=&(\sqrt{2}/a)\lbrace\Re (g_{k}(\eta) d^{R}_{k,l,m}) +i\Re
(g_{k}(\eta) d^{I}_{k,l,m})\rbrace\\
&=&(\sqrt{2}/a) |g_k(\eta)|\sqrt{\hbar} \lbrace  D^{R}_{k,l,m}
\cos(\alpha^R_{k,l,m} +\gamma_k) \\
&+&iD^{I}_{k,l,m}  \cos (\alpha^I_{k,l,m} +\gamma_k) \rbrace
\end{eqnarray}
so that writing $\alpha^{R,I}_{k,l,m} +\gamma = \alpha^{R,I}_{k,l,m} +\gamma^c_k +\Delta_k$ where $
\Delta_k=k(\eta-\eta^c_k)$, and using (\ref{Col2a}), we find
\begin{eqnarray}
\Gamma_{klm}(\eta) &=&(\sqrt{2}/a) |g_k(\eta)|\sqrt{\hbar} (1/2)
\lbrace
[x^R_{k,l,m,1}{\sqrt{1 +1/z_k^2}} Sin ( \Delta_k)\\
&+&
x^R_{k,l,m,2}( \cos (\Delta_k) + (1/z_k)  Sin( \Delta_k))]\\
&+& i[x^I_{k,l,m,1}{\sqrt{1 +1/z_k^2}}  Sin (\Delta_k)\\
&+& x^I_{k,l,m,2}( \cos (\Delta_k) + (1/z_k) Sin ( \Delta_k))]
\rbrace
\end{eqnarray}
Thus upon taking the ensemble average, and using $<<
x^A_{k,l,m,N}x^{A'}_{k',m,l,N'}>> = \delta_{k,k'}
\delta_{A,A'}\delta_{N,N'}$ (where $A, A'$ stand for $R$ or $I$
and $N,N'$ stand for $1$ or $2$) we find:
\begin{equation}
<<\Gamma_{klm}(\eta) \Gamma_{k'lm}(\eta) >> =\delta_{k,k'}
(\hbar/2a^2) |g_k(\eta)|^2 C(k)
\end{equation}
with $ C(k) $ given by (\ref{ExpCk}).  We thus have obtained a useful
expression for the desired quantity:
\begin{equation}
<<|\alpha_{lm}|^2>> =\sum_k\frac{s^2 U(k)^2}{(k^2+\mu)^2}(\hbar/2
a^2) |g_k(\eta)|^2 C(k)|f_{kl}(R_D)|^2
\end{equation}
Now we evaluate this sum.  To proceed we recall that the sum is
over the values of $k$ such that $kR=x^{(l)}_i$, the zeros of the
function $j_l(x)$.  Next we write $ f_{kl}(R_D)=c_{kl} j_l(k R_D)$
where $c_{kl}= k^{3/2} [\int_0^{x_i^{(l)}} (x j_l(x))^2)
dx]^{-1/2}$.  Thus $f_{k_il}(R_D)=c_{il} j_l(x^{(l)}_i R_D/R)$.
Now using eq 11.170 of \cite{Arfken} we find $ \int_0^{x_i^{(l)}}
(x j_l(x))^2) dx=(x_i^{(l)})^3j_{l+1}^2(x_i^{(l)})/2 $. Now the
approximate expression ( valid for large $x$) for the spherical
Bessel functions $j_l(x)= (1/x)\sin(x-l\pi/2)$ ( eq. 11.161a of
\cite{ Arfken}) we see that the $i^{th}$ zero of $j_l(x)$ is
$x_i^{(l)}= (i + l/2) \pi$, thus
$(j_{l+1}(x_i^{(l)})=(x_i^{(l)})^{-1}= [(i + l/2) \pi]^{-1}$.
Therefore $c_{kl}= k^{3/2} [(x_i^{(l)})/2]^{-1/2} = (k\sqrt{2/R})
$ where in the last step we used the fact that the allowed values
of $k_i$ must satisfy $ k_iR=x_i^{(l)} $.  We also use the fact
that $|g_k(\eta)|^2 =k/2$, and thus we find
\begin{equation}
<<|\alpha_{lm}|^2>> =\sum_k\frac{s^2 U(k)^2}{(k^2+\mu)^2}(\hbar/2
a^2) (k/2)  (2 k^2/R) |j^2_l(kR_D)|^2 C(k)
\label{APPalpha-lm}
\end{equation}
which we write as
\begin{equation}
<<|\alpha_{lm}|^2>> = \frac{\hbar s^2}{2 a^2 R }\sum_k E(k), \qquad
E(k):=\frac{U(k)^2k^3}{(k^2+\mu)^2} |j^2_l(kR_D)|^2 C(k).
\end{equation}
Now we use the uniform continuity of the function $F$, choose
$\epsilon >0$ and find the $\delta >0$ such that for $|k-k' |< \delta$
the corresponding $ |E_{(l)}(k) -E_{(l)}(k') |<\epsilon$, and then
write
\begin{equation}
\sum_k E(k) =
\delta^{-1}\sum_{n=0}^\infty \sum_{k\in [n\delta,
(n+1)\delta]} E_{(l)}(n\delta) \delta
\end{equation}
Now the number of allowed values of $k=(1/R) x^{l)}_i = (i + l/2)
\pi/R$ in the interval $[n\delta, (n+1)\delta]$ is $N= R\delta/\pi$.
Then we can write
\begin{equation}
\sum_k E(k)= \delta^{-1}\sum_{n=0}^\infty N
E_{(l)}(n\delta) \delta= (R/\pi) \int_0^\infty E_{(l)}(k)dk
\end{equation}
where in the last step we have taken the limit $\epsilon\to 0 $.  Thus
we finally obtain,
\begin{eqnarray}
<<|\alpha_{lm}|^2>>& =& \frac{\hbar s^2}{ 2 a^2 R }(R/\pi) \int_0^\infty
E_{(l)}(k)dk\\
&=&\frac{\hbar s^2}{ 2 a^2 \pi } \int_0^\infty  \frac{ U(k)^2
k^3}{(k^2+\mu)^2}   |j^2_l(kR_D)|^2 C(k)   dk
\end{eqnarray}
which is the same result that was obtained in the analysis of
Section \ref{sec_observation}.

%A note about dimensions:
% Recall $[s]= [LM]^{-1/2}$, $[\Psi]=1$, thus $[\Gamma]=[M]^{1/2}
%[L]^{-3/2}$.  On the other hand $
%[f_{kl}(r)]=[c_{kl}] = [L]^{-3/2}$. Therefore
%$[\Gamma_{klm}]=[M]^{1/2}$, and $[w_k]= [M]^{1/2}$.
% This is a  useful test  to check that we have not lost factors
%somewhere.

%-------------------------------------------------------------------
\section{On the  meaning of the Newtonian potential}
\label{app_newton}
%-------------------------------------------------------------------

Let us consider the space-time given by the perturbed metric:
\begin{equation}
ds^2=a(\eta)^2\left[-(1+ 2 \Psi) d\eta^2 +
(1-2\Psi)\delta_{ij} dx^idx^j\right]
\end{equation}
In this space-time light will travel according to
\begin{equation}
d\eta =\pm\left[\frac{(1- 2 \Psi)}{(1+ 2
\Psi)}\right]^{1/2}dx \approx \pm(1-2\Psi) dx
\end{equation}
We want to consider light signals being emitted from the surface of
last scattering at a given $\eta_D$ and arriving to us from all
angular directions. We will assume that we and the CMB radiation
emitting plasma are at rest in the background space-time so that both
we and the plasma fluid follow world lines of fixed $x$.

Thus a light signal that starts at $\eta_e^{(1)}$ at a given position
$x_e$ and arrives to the origin at $x=0$ at $\eta_o^{(1)}$, will
satisfy
\begin{equation}
\eta_o^{(1)} - \eta_e^{(1)} =
\int_0^{x_e}  (1-2 \Psi) dx
\end{equation}
Consider a second signal emitted an extremely short time after the
first one (we are thinking of fractions of a second), and from the
given position $x_e$ at $\eta_e^{(2)}= \eta_e^{(1)} + \delta
\eta_e$ and its arrival to the origin at $x=0$ at $\eta_o^{(1)}
=\eta_o^{(1)}+ \delta \eta_o$. It is clear then that
\begin{equation}
\eta_o^{(2)}-\eta_e^{(2)} =\int_0^{x_e}  (1-2 \Psi) dx
\end{equation}
and therefore $\delta \eta_e = \delta \eta_o$.  Now let us
consider the actual proper times measured by observers at rest
between the two emission events $\delta\tau_e$ and between the two
detection events $\delta\tau_o$. It is clear that $\delta\tau =
a(\eta)[1+2\Psi]^(1/2)\delta\eta \approx
a(\eta)[1+\Psi]\delta\eta$ so
\begin{equation}
\delta\tau_e =a(\eta_e)[1+\Psi_e]\delta\eta_e,  \qquad
\delta\tau_o =a(\eta_o)[1+\Psi_e]\delta\eta_0
\end{equation}
Thus we find
\begin{equation}
\frac {\delta\tau_o} {\delta\tau_e}=\frac {a(\eta_o)}
{a(\eta_e)} (1-\Psi_o)[1+\Psi_e]
\end{equation}
where it is clear that all the angular dependence is in the factor
$[1+\Psi_e]$, while $\frac {a(\eta_o)} {a(\eta_e)} (1+\Psi_o)$
represents the combine effect of the overall cosmic expansion and
the gravitational potential at our location. The quantity
$\frac{\delta\tau_o} {\delta\tau_e}$ is clearly encoding the
standard gravitational red-shift and is equal to the frequency
ratio $\nu_o /\nu_e$ which in turn is associated with the
temperature of the observed CMB. Note that the whole angular
dependence is in $\Psi_e$, so it is clear that what we have been
computing is indeed an observable quantity, and in that sense it
is gauge invariant.

%-------------------------------------------------------------------
\section{Alternative Collapse Scheme}
\label{app_alternative}
%-------------------------------------------------------------------

Here we want to consider an additional natural possibility for the
collapse.  The main motivation is to explicitly show how
observational data could be use to shed light into the details of
the collapse mechanism. The idea we consider here is that as it is
only the field's momentum which acts as a source, at leading
order, for the Newtonian potential, it should be only this
quantity that would be subjected to a change in the expectation
value during the collapse. This view seems to be close in spirit
to the ideas of Pen rose regarding the quantum uncertainties that
the gravitational potential would be inheriting from the matter
fields quantum uncertainties, as fundamental factors triggering
the collapse. In such a situation we would have a collapsed state
for which for $|\Theta\rangle$ after collapse:
\begin{equation}
\expec{\y_k^{R,I}(\eta^c_k)}_\Theta=0,\qquad
\expec{\pyRI_k(\eta^c_k)}_\Theta=X^{R,I}_{k}
\end{equation}
where $X^{R,I}_{k}$ are random variables, distributed according to
a Gaussian distribution centered at zero with spread
$\fluc{\pyRI_k}_0$, respectively. Thus
\begin{eqnarray}
%\expec{\y_k^{R,I}(\eta^c_k)}_\Theta&=0\\
\expec{\pyRI_k(\eta^c_k)}_\Theta&=&x^{R,I}_{k}\sqrt{\fluc{\py_k}_0}=x^{R
,I}_{k}|g_k(\eta^c_k)|\sqrt{\hbar L^3},
\end{eqnarray}
where $x_{k}$ are now distributed according to a Gaussian
distribution centered at zero with spread one. In this situation the
whole analysis of Section \ref{sec_observation} goes through, with
the only difference being that now we have:
\begin{equation}
F(k)=(x^R_k +ix^I_k) [\cos(\Delta_k) -(1/z_k)\sin
(\Delta_k)]
\end{equation}
so that
\begin{equation}
\langle F(\vec
k)\cc{F(\vec{k'})}\rangle =  \delta_{k,k'} [I +( 1- (1/z_k^2))
\sin^2(\Delta_k)  -(1/z_k)\sin (2\Delta_k)]
\end{equation}
And therefore
\begin{equation}
|\alpha_{lm}|^2_{Most Likely}=\frac{s^2   \hbar}{2 \pi a^2} \int
\frac{ U^2(x/R_D) C'(x/R_D)}{(x^2+\mu R_D^2)^2}    j^2_l(x) x^3 dx
\end{equation}
where now the function $C$ is slightly different from the one found in
the democratic collapse model, and is given by
\begin{equation}
C'(k)= [1 + \sin^2(\Delta_k) ( 1- (1/z_k^2)) -(1/z_k)\sin
(2\Delta_k)]
\end{equation}
We see that in principle the observations could help distinguish
between the different collapse models, and therefore it is clear
that the question of the exact mechanism for the origin of the
primordial fluctuations is both affected and can help shed light
on a fundamental issue in our understanding of quantum mechanics
as applied to the universe as a whole.

\end{appendix}
%-------------------------------------------------------------------


\begin{thebibliography}{99}

\bibitem{Boomerang-Maxima-WMAP}  ``Cosmological parameters From First
 results  of Boomerang"
Boomerang Collaboration (A.E. Lange et al.)
{\it  Phys. Rev. D},{\bf  63}, 042001,(2001);
 G. Hinshaw et. al., {\it Astrophys. J. Supp. }, {\bf 148}, 135
 (2003);
``Power Spectrum of Primordial Inhomogeneity Determined
 from four Year COBE DMR SKY Maps",
 K.M. Gorski , A.J. Banday , C.L. Bennett(NASA, Goddard), G. Hinshaw, A. Kogut , G.F. Smoot
, E.L. Wright ,   {\it Astrophys. J.} {\bf 464}, L11, (1996);
 ``First Year Wilkinson Micorowave Anisotropy Probe (WMAP) Observations: Preliminary Results"
C.L. Bennett et al. {\it   Astrophys. J. Suppl.} {\bf 148}, 1,(2003);
``First Year Wilkinson Micorwave Anisotropy Probe (WMAP)
Observations: Foreground Emission", C. Bennett et al. {\it Astrophys.
J. Suppl.} {\bf 148}, 97,(2003).

\bibitem{Guth} ``Quantum Mechanics of the scalar field in the new
  inflationary Universe", A. Guth and S.-Y. Pi {\it{ Phys.  Rev.  D}} {\bf 32}, 1899, (1985).

\bibitem{Hawking} ``Fluctuations in the Inflationary Universe",
S. W. Hawking {\it{Nucl. Phys.}} {\bf B 224}, 180,  (1983).

\bibitem{MotivInflation} See for instance discussion
 in ``The Early Universe", E.W. Kolb and M.S.
  Turner, Frontiers in Physics Lecture Note Series (Addison Wesley
  Publishing Company 1990).

\bibitem{Inflation-Wald} ``An Alternative to Inflation", S.
  Hollands and R.M. Wald,
  {\it Gen.Rel.Grav.}{ \bf 34}, 2043,(2002) [arXive: gr-qc/0205058].

\bibitem{Low-l-anomaly} ``Is the Universe out of Tune", G.D. Starkman and D. J. Schwartz,
{\it Scientific  American},  Vol 293, No 2,
48, (2005).

\bibitem{Starobinski1} ``Semiclassicality and decoherence of
  Cosmological perturbations", D. Polarski and A.A.
  Starobinsky,{\it Class.\ Quant.\ Grav.\ }  {\bf 13}, 377 (1996)
  [arXive: gr-qc/9504030]

\bibitem{Kiefer1} ``Origin of Classical Structure From Inflation", C. Kiefer
{\it Nucl.\ Phys.\ Proc.\ Suppl.\ } {\bf 88}, 255 (2000)
[arXiv:astro-ph/0006252], and references therein.

\bibitem{Parentani}  ``Space-time correlations in Inflationary Spectra",
D. del Campo, and R. Parentani, {\it Phys.\ Rev.\ D} {\bf 70}, 105020
(2004) [arXiv:gr-qc/0312055]

\bibitem{Zurek} ``Environment Induced Superselection In Cosmology",
 W.H. Zurek, in {\it Moscow 1990, Proceedings, Quantum gravity} (QC178:S4:1990), 456-472.
 (see High Energy Physics Index 30 (1992) No. 624)

%\bibitem{Starobinski2}.....

\bibitem{Decoherence} ``Decoherence Functional and Inhomogeneities in
  the Early Universe", R.  Laflamme and A. Matacz,
{\it Int.\ J.\ Mod.\ Phys.\ D } {\bf 2}, 171 (1993)
[arXiv:gr-qc/9303036]

\bibitem{branderberger} ``Lectures on the Theory of Cosmological
  Perturbations", R.H.  Brandenberger,{\it Lect.\ Notes Phys.\ }  {\bf 646}, 127 (2004)
  [arXiv:hep-th/0306071]

\bibitem{Mukhanov} ``Gauge Invariant Cosmological Perturbations" R.
  Branderberger H. Feldman and V. Mukhavov, Preprint Brown HET 845
  (1992); Phys. Rep. { \bf 215}, 203, (1992)

\bibitem{Padmanaban} ``Cosmology and Astrophysics Through Problems", T.
  Padmanabhan (Cambridge University Press 1996).

\bibitem{Liddle} `` Cosmological Inflation and Large Scale Structure",
  A.R. Liddle and D.H. Lyth (Cambridge University Press
  2000).

\bibitem{Castagnino:2002ic} ``The self-induced approach to decoherence in cosmology,''
  M. Castagnino and O. Lombardi,{\it  Int. J. Theor. Phys.} {\bf 42}, 1281 (2003)
  [arXiv:quant-ph/0211163].


\bibitem{Halliwell-85} ``Origin of Structure in the Universe" J.J. Halliwell and S. W. Hawking,
{\it Phys. Rev. D}, {\bf 31}, 1777,(1985).


\bibitem{Correlations}``Inflationary Cosmological Perturbations of Quantum Mechanical Origin"
J. Martin, {\it Lect.\ Notes Phys.\ } {\bf 669}, 199 (2005)
  [arXiv:hep-th/0406011];
 `` Best Unbiased Estimates for Microwave background Anisotropies", L.P. Grishchuk and J. Martin,
 {\it Phys.\ Rev.\ D} {\bf 56}, 1924 (1997)
  [arXiv:gr-qc/9702018]

\bibitem{Penrose} ``The Emperor's New Mind", R. Penrose ( Oxford
  University Press 1989).

\bibitem{COBE} ``Structure in the COBE DMR first year maps"
 G.F. Smoot {\it et. al., Astrophys. J.}, {\bf 396}, L1 (1992).

\bibitem{Adler} S. Adler, (Princeton) Private Communication.

\bibitem{Nonlocality} ``Nonlocality in Quantum Physics",
 A. Anaoljevich Grib and W. Alves Rodrigues Jr. (Kluwer Academic / Plenum Publishers 1999).

\bibitem{Hartle-1997} ``Quantum Cosmology  Problems for the 21${}^{st}$ Century", J.~B.~Hartle, arXive:
gr-qc/9701022.

\bibitem{Hartle-2005} ``Generalized Quantum mechanics for Quantum Gravity",J.~B.~Hartle, arXive: gr-qc/0510126.

\bibitem{Hartle-1993}``The Reduction of the State Vector and Limitations on Measurement
in Quantum Mechanics of Closed Systems", J.~B.~Hartle, in
``Directions in Relativity. Vol. 2: Proceedings", B.L. Hu and T.A.
Jacobson (eds.), Cambridge University Press, Cambridge, 1993.
[arXive: gr-qc/9301011]

\bibitem{kent} ``Consistent Sets Yield Contrary Inferences in Quantum
Theory", A. Kent, Phys. Rev. Lett. 78, 28749 (1997);  ``Comments on
``Consistent Sets Yield Contrary Inferences in Quantum Theory"", R.
Griffiths and J.B. Hartle,Phys.\ Rev.\ Lett.\  {\bf 81}, 1981 (1998)
[arXiv:gr-qc/9710025]

\bibitem{Halliwel-89}``Decoherence in Quantum Cosmology", J.J. Halliwell,
{\it Phys. Rev. D}, {\bf 39}, 2912,(1989)


%\bibitem{Starobinski1Sec} See Section 3 of Ref. \cite{Starobinski1}.
%% ``Semiclasicallity and decoherence of
%% Cosmological perturbations", David Polarski and Alexei A. Starobinsky, arXive: gr-qc/9504030 (1996).
%
%\bibitem{ParentaniSec} See Section V of Ref. \cite{Parentani}.
%% ``Space-time correlations in Inflationary Spectra: a wave packet analysis",
%% D. del Campo, and R. Parentani, arXive: gr-qc/0312055, Jul 2004.

\bibitem{Hu} ``Recent Advances in Stochastic Gravity Theory and
  Issues", B. L.  Hu and E. Verdaguer, arXive: gr-qc/0110092.

\bibitem{More-Decoherence}
  ``Decoherence in Quantum Cosmology at the onset of Inflation", A.O. Barvinsky, A.Y. Kamenshchik, C. Kiefer, and
I.V. Mishakov, {\it Nucl.\ Phys.\ B} {\bf 551}, 374 (1999)
  [arXiv:gr-qc/9812043]

\bibitem{Bell} ``Speakable and Unspeakable in Quantum Mechanics", J.
  S. Bell (Cambridge University Press 1987).

\bibitem{Mermin} ``Is the Moon There when nobody Looks?"
 D. Mermin {{\it Physics  Today }} {\bf 32}, 38, (1985).

\bibitem{Aspect} ``Experimental realization of Einstein-Podolsky-Rosen-Bohm Gedankenexperiment:
A New violation of Bell's inequalities"
 A. Aspect, P. Grangier, G. Roger,
{\it Phys. Rev. Lett.} {\bf 49}, 91, (1982).

\bibitem{GRW} ``A Unified Dynamics For Micro And Macro Systems",
G. C. Ghirardi, A.  Rimini, and T.  Weber,{\it  Phys. Rev.} {\bf D 34}, 470, (1986).

%\bibitem{deviation} ``Power Spectrum of Primordial Inhomogeneity Determined
% from four Year COBE DMR SKY Maps",
% K.M. Gorski , A.J. Banday , C.L. Bennett(NASA, Goddard), G. Hinshaw, A. Kogut , G.F. Smoot
%, E.L. Wright ,   {\it Astrophys. J.} {\bf 464}, L11, (1996);
% ``First Year Wilkinson Micorowave Anisotropy Probe (WMAP) Observations: Preliminary Results"
%C.L. Bennett et al. {\it   Astrophys. J. Suppl.} {\bf 148}, 1,(2003);
%``First Year Wilkinson Micorwave Anisotropy Probe (WMAP)
%Observations: Foreground Emission", C. Bennett et al. {\it Astrophys.
%J. Suppl.} {\bf 148}, 97,(2003).

\bibitem{Asher} Quantum Theory: Concepts and Methods, A. Peres (Kluwer, Academic Publishers, 1993).

\bibitem{Arfken} Mathematical Methods  for Physicists,  G. Arfken,
 (Academic Press New York 1970).

\bibitem{Lombardo:2005iz} ``Decoherence during inflation: The generation of classical
  inhomogeneities,''
  F.~C.~Lombardo and D.~Lopez Nacir,
 Phys.\ Rev.\ D {\bf 72}, 063506 (2005)
  [arXiv:gr-qc/0506051].

\end{thebibliography}
\end{document}